\documentclass[a4paper,oneside,11pt,notitlepage]{report}

\usepackage{geometry}
\usepackage[english]{babel} 
\usepackage[T1]{fontenc}
\usepackage[ansinew]{inputenc}
\usepackage{lmodern} 
\usepackage{eurosym}
\usepackage{cancel}
\usepackage{nameref} 

\usepackage{graphicx} 
\usepackage[position=top]{subfig}   
\usepackage{caption}
\usepackage{amsmath}
\usepackage{amsthm}
\usepackage{amsfonts}
\usepackage{dsfont}
\usepackage{xcolor}
\usepackage{mathabx}
\usepackage{array}
\usepackage{slashbox}
\usepackage{mathrsfs}
\usepackage{bm}
\usepackage{lscape}
\usepackage{verbatim}
\usepackage{multicol}
\usepackage{multirow}
\usepackage{changepage}
\usepackage{footnote}
\usepackage{makecell}
\usepackage{lscape}
\usepackage[symbol*]{footmisc}
\usepackage{lipsum}
\usepackage[round]{natbib}
\usepackage{breqn}
\usepackage[bookmarks=false]{hyperref}

\newcommand\independent{\protect\mathpalette{\protect\independenT}{\perp}}
\def\independenT#1#2{\mathrel{\rlap{$#1#2$}\mkern2mu{#1#2}}}

\newcommand{\bx}{{\bm x}}
\newcommand{\bX}{{\bm X}}
\newcommand{\btheta}{\bm\theta}

\newcommand{\bu}{{\bm u}}
\newcommand{\bF}{{\bm F}}

\newcommand{\E}{{\mathbb{E}}}
\newcommand{\V}{{\mathbb{V}\text{ar}}}
\newcommand{\PP}{{\mathbb{P}}}
\newcommand{\RR}{\mathbb{R}}
\newcommand{\ind}{\mathds{1}}

\definecolor{dgreen}{rgb}{0,0.5,0}
\definecolor{dblue}{rgb}{0,0,0.9}
\definecolor{dred}{rgb}{0.7,0.0,0.1}
\definecolor{dgold}{rgb}{0.5,0.3,0.0}
\definecolor{dvio}{rgb}{0.6,0.3,0.5}
\definecolor{gray}{rgb}{0.5,0.5,0.5}
\definecolor{myblue}{rgb}{0.137,0.466,0.741}

\DefineFNsymbolsTM{myfnsymbols}{
  \textasteriskcentered *
  \textdagger    \dagger
}
\numberwithin{equation}{section} 
\theoremstyle{plain}
\newtheorem{theorem}{Theorem}[section] 			 
\newtheorem{corr}[theorem]{Corollary}     

	\makeatletter
	\def\namedlabel#1#2{\begingroup
			#2%
			\def\@currentlabel{#2}%
			\phantomsection\label{#1}\endgroup
	}
	\makeatother 

\hypersetup{colorlinks,citecolor=blue}

\title{Semiparametric Copula Quantile Regression\\ for Complete or Censored Data}
\author{Micka\"{e}l De Backer, Anouar El Ghouch, Ingrid Van Keilegom
					\footnote{\small Universit\'{e} catholique de Louvain, Institut de Statistique, Biostatistique et Sciences Actuarielles. Voie du Roman Pays 20, B-1348 Louvain-la-Neuve, Belgium. E-mail: mickael.debacker@uclouvain.be.}
									\\
									\\
				\normalsize{\textit{Universit\'{e} catholique de Louvain}}\\
}

\begin{document}


\maketitle
\begin{abstract}
When facing multivariate covariates, general semiparametric regression techniques come at hand to propose flexible models that are unexposed to the curse of dimensionality. In this work a semiparametric copula-based estimator for conditional quantiles is investigated for complete or right-censored data. In spirit, the methodology is extending the recent work of \citet{NEGB13} and \citet{NEGVK15}, as the main idea consists in appropriately defining the quantile regression in terms of a multivariate copula and marginal distributions. Prior estimation of the latter and simple plug-in lead to an easily implementable estimator expressed, for both contexts with or without censoring, as a weighted quantile of the observed response variable. In addition, and contrary to the initial suggestion in the literature, a semiparametric estimation scheme for the multivariate copula density is studied, motivated by the possible shortcomings of a purely parametric approach and driven by the regression context. The resulting quantile regression estimator has the valuable property of being automatically monotonic across quantile levels, and asymptotic normality for both complete and censored data is obtained under classical regularity conditions. Finally, numerical examples as well as a real data application are used to illustrate the validity and finite sample performance of the proposed procedure. 
\end{abstract}

\newpage
\section{Introduction}

Quantile regression is a prevailing method when it comes to investigating the possible relationships between a $d$-dimensional covariate $\bX$ and a response variable $T$. Since the seminal work of \citet{KB78}, quantile regression has received notable interest in the literature on theoretical and applied statistics as a very attractive alternative to the classical mean regression model based on quadratic loss. As the latter only captures the central tendancy of the data, there are many cases and nice examples where mean regression is uninformative with respect to studying the conditional upper or lower quantiles. For an interesting application, see for example \citet{EKJ08}. A comprehensive review of quantile regression as a robust (to outliers) and flexible (to error distribution) method can be found in \citet{K05}.

A wide literature on the estimation of the quantile regression function is devoted to the case where the response variable $T$ is completely observed. This is not necessarily the case in survival analysis, where right censoring of $T$ may arise, i.e. instead of fully observing the variable of interest, one only observes the minimum of it and a censoring variable. For instance, in clinical studies, censoring may occur because of the withdrawal of patients from the study, the end of the follow-up period, etc. In this context, quantile regression becomes attractive as an alternative to popular regression techniques like the Cox proportional hazards model or the accelerated failure time model, as is argued in \citet{KB01}, \citet{KG01} and \citet{P03}. Additional appealing properties of the method include the fact that it allows for modelling heterogeneity of the variance and it does not necessarily impose a proportional effect of the covariates on the hazard over the duration time as opposed to the popular Cox model.

As is the case for the uncensored situation, existing literature on censored quantile regression includes fully parametric, semiparametric and nonparametric methodologies. When several covariates are to be taken into account, fully parametric methodologies are known to be highly sensitive to model misspecification and may lack the flexibility needed for an adequate modelling. On the other hand, in spite of their great flexibility, fully nonparametric methods such as local linear smoothing proposed by \citet{EGVK09} are typically affected by the curse of dimensionality. In light of these restrictions,  semiparametric estimation procedures such as a single-index model suggested by \citet{BEGVK14} come at hand when the dimension of the covariate is high. It is the object of this paper to extend the rather sparse literature on flexible multidimensional methodologies in the context of censored quantile regression.

Censored quantile regression was first introduced by \citet{P86} for linear models and fixed censoring, that is, presuming that the censoring times are the same for all observations. For random censoring, \citet{YJW95} proposed a semiparametric linear median regression model, assuming that the survival time and the censoring variable are unconditionally independent. Despite this important contribution, the proposed procedure involves solving non-monotone discontinuous equations, hereby introducing practical and computational difficulties. Furthermore, the unconditional independence assumption may sound restrictive as conditional independence, given the covariates, seems more natural in certain applications.

Based on conditional independance between the survival time and the censoring variable, \citet{P03} developed a novel estimating procedure based on the idea of redistribution-of-mass introduced by \citet{E67}. A major shortcoming of it is the need for a global linear assumption, that is, in order to estimate the $\tau$-th conditional quantile, one needs to assume the linearity of all the conditional functionals at lower quantiles. Extending \citeauthor{P03}'s work, \citet{WW09} relaxed this constraint by only assuming linearity at one pre-specified quantile level of interest. Recently, \citet{LT13} proposed an alternative to \citeauthor{WW09}'s methodology by extending the work of \citeauthor{YJW95} in order to relax the constraining unconditional independance assumption and provide an efficient algorithm for estimation. Still, a linear approach may be too restrictive for real data applications. 


Finally, an interesting alternative approach to the above-mentioned literature was proposed by \citet{BEGVK14}, where a single-index model for the conditional quantile function is studied under the assumption of independence between the covariates and the censoring variable. The single-index structure assumes that the objective function depends linearly on the covariates through an unknown link function, making the proposed model (i.e. under the aforementioned assumption) insensitive to the curse of dimensionality since the nonparametric part is of dimension one.

In this paper, we aim to extend the literature on multivariate quantile regression estimation in the possible presence of censored data by providing a rich, flexible and robust alternative based on the copula function that defines the dependence structure between the variables of interest. By taking advantage of copula modelling, we intend to provide a new class of estimators that would allow practitioners to analyse, in a flexible way, multidimensional survival data. Actually, our methodology is, in essence, an extension of the recent work of \citet{NEGB13} and \citet{NEGVK15}, as the central idea is to express the conditional quantile function in terms of an appropriate copula density and marginal distributions. In their original paper, \citet{NEGB13} suggested subsequently to leave the marginal distributions unspecified while assuming a parametric model for the copula. Overall, their suggested approach results in a semiparametric regression estimator that is not exposed to the curse of dimensionality. However, in order to avoid possible shortcomings highlighted by \citet{DVHV14} that are induced by the misspecification of the parametric copula, we propose in this work, both for complete and censored data, an alternative semiparametric estimation strategy for the copula itself. Our resulting methodology is flexible for multidimensional data with or without censoring, easy to implement and does not require any iterative procedure in opposition to existing semiparametric alternatives. 

The rest of this paper is organised as follows. Developing the copula-based estimation procedure for the quantile regression is the topic of Section \ref{section:methodology}. The asymptotic properties of the proposed estimator are obtained in Section \ref{section:properties} and the finite sample performance is illustred by means of Monte Carlo simulations in Section \ref{section:simulations}, where both the semiparametric copula estimation strategy and the overall performance of our estimator are investigated. Section \ref{section:application} provides a brief application to real data. Lastly, the proofs of our asymptotic properties are deferred to the Appendix.

\section{Methodology and Estimation}\label{section:methodology}

\subsection{Copula-based estimator for complete data}\label{sect:background}
Let $\bX = (X_1,\ldots,X_d)^{\mathsf{T}}$ be a covariate vector of dimension $d \geq 1$ and $T$ be a (time-to-event) response variable with marginal continuous cumulative distribution functions (c.d.f.) $F_1,\ldots,F_d$ and $F_T$, respectively. Throughout this paper, we denote by $f_j$ and $f_T$ the density of $X_j, j=1,\ldots,d,$ and $T$, respectively. From the pioneering work of \citet{S59}, for a given $\bx = (x_1,\ldots,x_d)^{\mathsf{T}}$, the c.d.f. of $(T,\bX)$ evaluated at $(t,\bx)$ can be expressed as $C_{T\bX}(F_T(t),\bF(\bx))$, where $\bF(\bx) = (F_1(x_1),\ldots,F_d(x_d))^{\mathsf{T}}$ and $C_{T\bX}$ is the unique copula distribution of $(T,\bX)$ defined by $C_{T\bX}(u_0,u_1,$ $\ldots,u_d) = \PP(U_0 \leq u_0, U_1 \leq u_1, \ldots, U_d \leq u_d)$, with $U_0 = F_T(T)$ and $U_j = F_j(X_j)$, $j=1,\ldots,d$. From Sklar's theorem, it is clear that the copula $C_{T\bX}$ disjoints the marginal behaviours of $T$ and $\bX$ from their dependence structure, hence allowing a great modelling flexibility. For a book length treatment of copulas, see \citet{N06} and \citet{J14}. 

The object of interest of this paper, the $\tau$-th conditional quantile function of the dependent variable $T$ given $\bX=\bx$, denoted by $m_{\tau}(\bx)$, is defined for any $\tau \in (0,1)$ as $m_{\tau}(\bx) = \inf \{t: F_{T|\bX}(t|\bx) \geq \tau \}$ where $F_{T|\bX}$ is the conditional c.d.f. of $T$ given $\bX$, or, equivalently,
\begin{eqnarray}\label{eq:QR}
m_{\tau}(\bx) = \arg\min_a  \E\big(\rho_{\tau}(T-a)|\bX=\bx\big),
\end{eqnarray}
where $\rho_{\tau}(u) = u(\tau - \ind(u<0))$ is the so-called ``check'' function, and $\ind(\cdot)$ is the indicator function. 

To motivate our approach, let us suppose, for the moment, that there is no censoring and that we observe an i.i.d. sample $(T_i,\bX_i), i=1,\ldots,n,$ from $(T,\bX)$. In this context, following the definition of a copula function, \citet{NEGVK15} noted that the conditional quantile function of $T$ given $\bX=\bx$ may be expressed as
\begin{eqnarray}\label{eq:CQR}
m_{\tau}(\bx) = \arg\min_a  \E\Big[\rho_{\tau}(T-a)\,c_{T\bX}(F_T(T),\bF(\bx))\Big],
\end{eqnarray}
where $c_{T\bX}(u_0,\bu) \equiv c_{T\bX}(u_0,u_1,\ldots,u_d) = \partial^{d+1}C_{T\bX}(u_0,u_1,\ldots,u_d)/\partial u_0 \partial u_1 \ldots \partial u_d$ is the copula density corresponding to $C_{T\bX}$. Consequently, any given estimators $\widehat{c}_{T\bX}$, $\widehat{F}_T$ and $\widehat{F}_j$ of $c_{T\bX}$, $F_T$ and $F_j$, $j=1,\ldots,d$, respectively, automatically yield an estimator of $m_\tau(\bx)$ given by 
\begin{eqnarray}\label{eq:CQR estimator}
\widehat{m}_{\tau}(\bx) = \arg\min_a  \sum_{i=1}^n\rho_{\tau}(T_i-a)\,\widehat{c}_{T\bX}(\widehat{F}_T(T_i),\widehat{\bF}(\bx)),
\end{eqnarray} 
with $\widehat{\bF}(\bx) = (\widehat{F}_1(x_1),\ldots,\widehat{F}_d(x_d))^{\mathsf{T}}$. As indicated earlier, \citeauthor{NEGVK15} suggest to estimate the marginals nonparametrically and to consider a parametrization of the copula density, that is, assume that the latter belongs to a  certain parametric family of copula densities $\mathcal{C} = \{ c(u_0,\bu;\btheta), \btheta\in \Theta \subset \RR^p\}$. 

\begin{figure}[t!]
  		\captionsetup[subfloat]{captionskip=0pt,position=below}
     \centering
     \subfloat[][]{\includegraphics[width=0.45\linewidth,height=3.8cm]{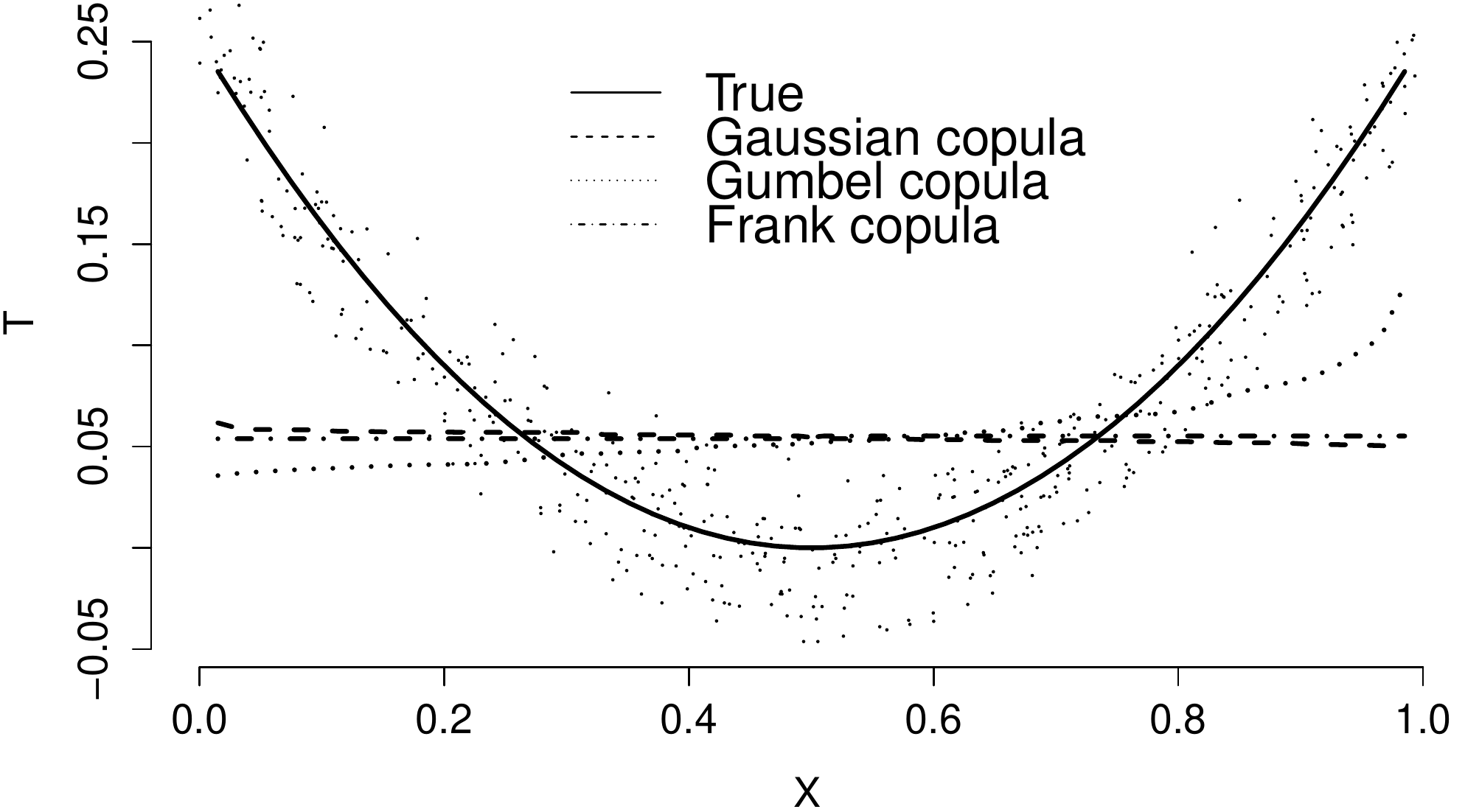}\label{fig:Ex_Dette_para}}
     \hspace{0.5cm}
		 \subfloat[][]{\includegraphics[width=0.45\linewidth,height=3.8cm]{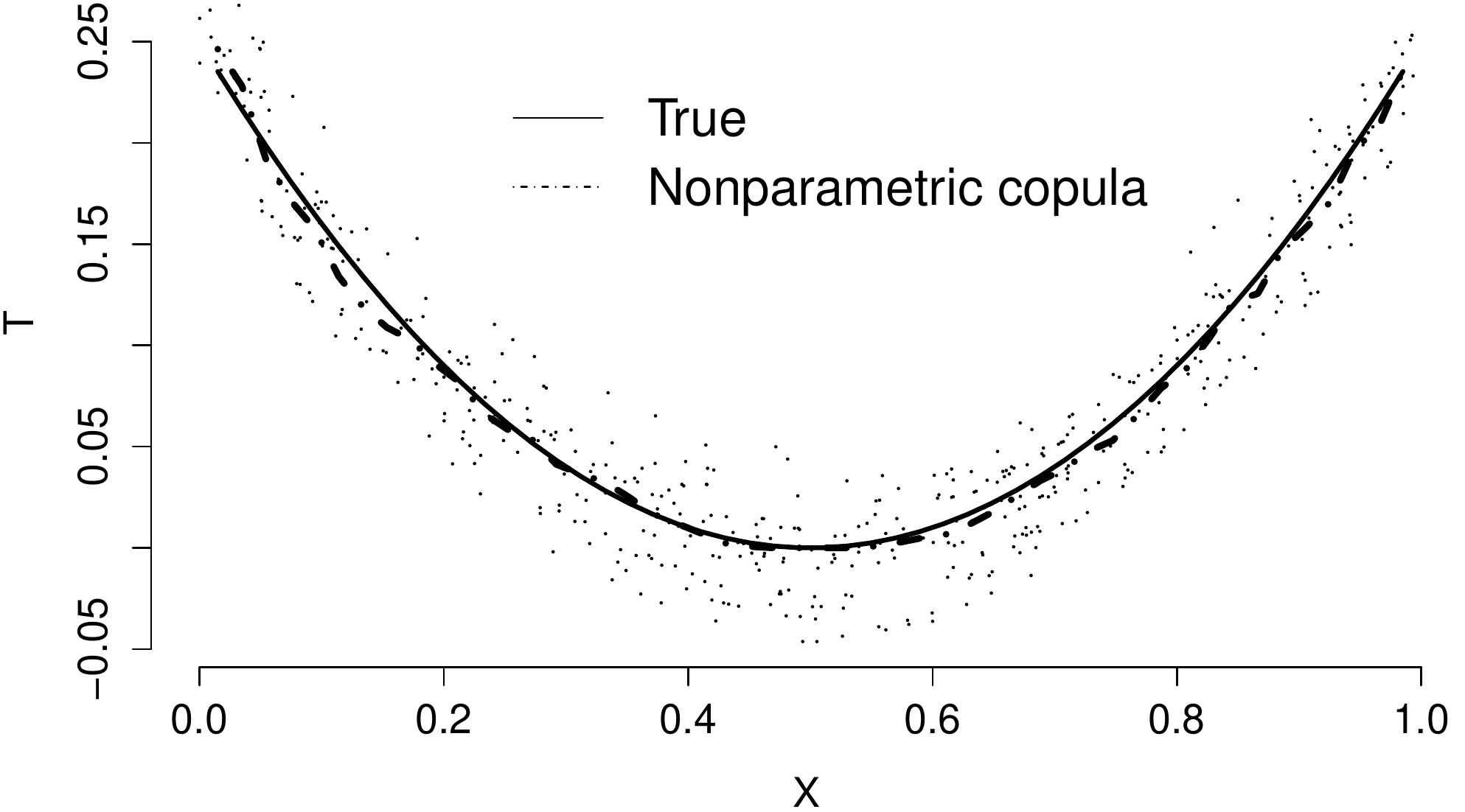}\label{fig:Ex_Dette_nonpara}}
     \caption{Copula-based quantile regression estimates of the minimalistic example. The Gaussian, Gumbel and Frank copulas are used for parametric copula estimation in (a), while (b) depicts the regression fit resulting from a nonparametric estimation of the copula density using the procedure of \citet{GCP14} (more details given below).}
     \label{fig:Ex_Dette}
\end{figure}

In this paper however, we propose a novel semiparametric strategy for the estimation of the copula density, motivated by the issues related to the possible misspecification of the parametric approach. To highlight this shortcoming and illustrate how one may circumvent it, we consider the simplistic example reported by \citet{DVHV14} with a single covariate, where $(T_i,X_i), i=1,\ldots,n,$ are i.i.d. random variables with $T_i = (X_i-0.5)^2 + \sigma\epsilon_i$, $X_i \sim U[0,1]$, $\sigma = 0.025$ and $\epsilon_i, i=1,\ldots,n,$ are i.i.d. standard normal random variables. In this situation, where the true quantile regression function is non-monotonic in the covariate, it is found that most of the common parametric copula families still yield a monotone estimation of the regression function, thereby providing a rather poor fit of the latter. This is illustrated in Figure \ref{fig:Ex_Dette_para}, where the estimation is carried out for $\tau=0.5$, $n=500$ and using three common parametric copulas.

As the roots of the above-mentioned limitation are not intrinsic to a copula-based approach, but rather to be attributed to the limited set of parametric copula families existing in the literature, a natural alternative, for low dimensional covariates only, would be to consider a fully nonparametric estimation of the copula density itself. The resulting, and adequate, quantile regression estimation is depicted in Figure \ref{fig:Ex_Dette_nonpara}.

Recalling that we intend to handle multivariate covariates in this paper, we will not adopt a purely nonparametric approach, but rather prefer a copula estimation strategy that provides sufficient flexibility to the multidimensional estimator while dodging dimension related constraints. More specifically, we note that any multivariate copula density can be decomposed into two parts as follows: 
\begin{eqnarray}
c_{T\bX}\big(F_T(t), \bF(\bx)\big) &=& c_{TX_1}\big(F_T(t),F_1(x_1)\big) \times \ldots \times c_{TX_d}\big(F_T(t),F_d(x_d)\big)\label{eq:multivariate cop part 1}\\
&&\times c_{X_1\ldots X_d|T}\big(F_{1|T}(x_1|t),\ldots,F_{d|T}(x_d|t)|t\big), \label{eq:multivariate cop part 2}
\end{eqnarray}
where $F_{j|T}$, $j=1,\ldots,d$, denotes the conditional c.d.f. of $X_j$ given $T$. The first part of the decomposition contains the product of $d$ bivariate copula densities related to the dependence of $T$ with every covariate, whereas the second part captures the conditional dependence of $\bX$ given $T=t$. In the general regression context, part \eqref{eq:multivariate cop part 1} may then be interpreted as the dependence of actual interest since it focuses on the relationship between the response variable with every covariate. On the contrary, part \eqref{eq:multivariate cop part 2} may be viewed in such framework as a `noisy' dependence, or, more precisely, a correction parameter for possible (conditional) dependence among covariates. Consequently, a natural reasoning suggests to provide as much flexibility as possible to the modelling of part \eqref{eq:multivariate cop part 1}, while keeping the estimation of part \eqref{eq:multivariate cop part 2} uncomplicated. We therefore advocate to estimate nonparametrically the $d$ bivariate copulas of interest and, subsequently, exploit standard parametric techniques for the second part of the multivariate copula density.

Concerning the nonparametric estimation of bivariate copula densities, several methodologies have been proposed in the literature. To estimate for instance $c_{TX_1}$, a tempting, yet naive, approach would be to opt for standard multivariate kernel techniques. That is, given a bivariate sample $(U_{0i}, U_{1i}), i=1,\ldots,n$, from $(F_T(T),F_1(X_1))$, one would estimate the copula density $c_{TX_1}$ at $(u_0,u_1) \in [0,1]^2$ by 
\begin{eqnarray*}
\frac{1}{n \, |\bm{H}|^{1/2}} \sum_{i=1}^n \bm{K}\left( \bm{H}^{1/2} \begin{pmatrix} u_0 - U_{0i} \\[-3.5pt] u_1 - U_{1i} \end{pmatrix}\right),
\end{eqnarray*}
where $\bm{H}$ is a symmetric positive-definite bandwidth matrix and $\bm{K}:\RR^2 \rightarrow \RR$ is a bivariate kernel function. This technique is called naive in this context since it completely ignores the fact that the density of interest is only supported on the unit square. This boundedness property is at the origin of distinct schemes designed to correct for well-known bias issues at the boundaries, see for example the mirror reflection method (\citet{GM90}) or the boundary kernel method (\citet{CH07}). An appealing alternative was furthermore proposed by \citet{CFS06} and \citet{GCP14}, where the main idea is to appropriately project the initial data on an unbounded support with the purpose of then estimating the obtained bivariate transformed density by means of standard techniques (standard kernel (\citeauthor{CFS06}) or polynomial local-likelihood (\citeauthor{GCP14})). Using the invariance property of copulas to increasing transformations of their margins, the estimation of the copula density is then obtained by back-transformation on the unit square. That is, using the example of a probit transformation, one may estimate the copula density $c_{TX_1}$ at $(u_0,u_1) \in [0,1]^2$ by 
\begin{eqnarray*}
\frac{\widehat{f}_{01}\big(\Phi^{-1}(u_0),\Phi^{-1}(u_1)\big)}{\phi\big(\Phi^{-1}(u_0)\big)\phi\big(\Phi^{-1}(u_1)\big)},
\end{eqnarray*}
where $\phi$ and $\Phi$ stand for the standard normal density and c.d.f., respectively, and where $\widehat{f}_{01}$ is a bivariate density estimator of the projected data $(\Phi^{-1}(U_{0i}),\Phi^{-1}(U_{1i})), i=1,\ldots,n$. This transformation technique, coupled with polynomial local-likelihood estimation for $f_{01}$, in order to allow for possible unbounded copula density estimates, is shown to outperform its competitors in most scenarios in a detailed simulation study in \citeauthor{GCP14}. An exhaustive comparison of existing methodologies for bivariate estimation may be found in \citet{Thesis_Nagler}. Furthermore, fully nonparametric multidimensional copulas are studied in \citet{HHS12} and \citet{NC15}. 

Regarding the second part \eqref{eq:multivariate cop part 2} of the $(d+1)$-dimensional copula, standard high-dimensional parametric techniques involve, among others, nested Archimedean copulas (see e.g. \citet{HP13} and \citet{J14}), factor copulas (see e.g. \citet{OP12}) and, arguably the most popular, vine copulas (see e.g. \citet{C10}, \citet{J14} and references therein). Note, however, that for the estimation of \eqref{eq:multivariate cop part 2} in practice, one is first advised to adopt the so-called \textit{simplifying assumption} which stipulates that the conditioning on $T=t$ is fully captured by the conditional marginals. In other words, in \eqref{eq:multivariate cop part 2}, the conditional copula itself is not affected by the conditioning on $T$. This assumption turns out to be the cornerstone of vine copula models as it keeps them tractable for inference and model selection. For more details about this and its implications, see \citet{HHAF10} and \citet{SJC13}.
 
Conclusively, in this article we propose to adopt the following detailed procedure for the modelling and estimation of the multivariate copula density:
\begin{itemize}
	\item[\namedlabel{cop:step1}{(1)}] Based on original observations $(T_i,X_{1i},\ldots,X_{di}), i=1,\ldots,n,$ construct `pseudo-observations', needed for the estimation of \eqref{eq:multivariate cop part 1}, using rescaled versions of empirical distributions: 
	\begin{align*}
\widehat{U}_{0i} = \frac{1}{n+1}\sum_{k=1}^{n} \ind(T_k \leq T_i) && \widehat{U}_{ji} = \frac{1}{n+1}\sum_{k=1}^{n} \ind(X_{jk} \leq X_{ji}), i=1,\ldots,n, \quad j=1,\ldots,d, 
\end{align*}
where the factor $1/(n+1)$, commonly adopted in the copula literature, aims at keeping the constructed observations in the interior of $[0,1]$.
	\item[\namedlabel{cop:step2}{(2)}] Based on $(\widehat{U}_{0i},\widehat{U}_{ji}), i=1,\ldots,n,$ estimate each bivariate copula density $c_{TX_j},$ $j=1,\ldots,d,$ in \eqref{eq:multivariate cop part 1} using a bivariate kernel density estimator. This can be achieved via the local-polynomial probit methodology of \citeauthor{GCP14}, or any other estimator satisfying assumption \ref{cond:cop_hat} given below. 
	\item[\namedlabel{cop:step3}{(3)}] Compute the pseudo-observations needed for the estimation of \eqref{eq:multivariate cop part 2} as: $$ \widehat{F}_{j|T}(X_{ji}|T_i) = \int_0^{\widehat{U}_{ji}} \widehat{c}_{TX_j}(\widehat{U}_{0i},s) \mathrm ds.$$ This relationship is at the origin of the sequential nature of the vine copula estimation scheme (see e.g. \citet{C10}). 
	\item[\namedlabel{cop:step4}{(4)}] Lastly, for the estimation of $c_{X_1\ldots X_d|T}$, adopt the simplifying assumption and use standard parametric vine techniques on the dataset $(\widehat{F}_{1|T}(X_{1i}|T_i),\ldots,\widehat{F}_{d|T}(X_{di}|T_i))$, $i=1,\ldots,n$.
\end{itemize}

\subsection{Copula-based estimator for censored data}\label{subsect:our proc}

In the presence of censoring, the estimation equation \eqref{eq:CQR estimator} becomes inappropriate as we do not fully observe the response variables $T_i$. Instead, we only observe a sequence of i.i.d. triplets $(Y_i,\Delta_i,\bX_i)$, $i=1,\ldots,n,$ from $(Y,\Delta,\bX)$, where $Y = \min(T,C)$,  $\Delta = \ind(T \leq C)$ and $C$ denotes the censoring variable, assumed to be independent of $T$ given $\bX$. In order to take censoring into account in the estimation procedure, the first step is to note that, for any measurable function $\varphi: \RR \rightarrow \RR$,
\begin{eqnarray}\label{eq:cond expec}
\E\big(\varphi(T)|\bX=\bx\big) = \E \left(\varphi(Y) \frac{\Delta}{1-G_C(Y-|\bx)} \Big| \bX = \bx\right),
\end{eqnarray} 
where $G_C(c|\bx) = \PP(C \leq c|\bX=\bx)$ denotes the conditional distribution of $C$ given $\bX=\bx$. This, along with \eqref{eq:QR}, suggests a natural way to handle censoring for quantile regression by replacing the function $\varphi$ with the check function.

At this stage, a naive way of trying to take profit of copula modelling would be to consider introducing copulas in the obtained conditional expectation. Note, however, that the conditional expectation in \eqref{eq:cond expec} is in fact the joint conditional expectation of $(Y,\Delta)$ given $\bX=\bx$. Adopting an analogous reasoning as the one presented by \citeauthor{NEGVK15} for the uncensored case at this point would therefore result in the insertion of the joint copula of $(Y,\Delta,\bX)$, hence exposing the estimation procedure to the lack of uniqueness of the copula given that $\Delta$ is a discrete (binary) variable. In this situation, to quote \citet{Em09}, ``everything that \textit{can} go wrong, \textit{will} go wrong''. Details about copulas for discrete variables may be found in \citet{GN07}.   

\newpage
Instead, the idea is to work on the joint conditional expectation so as to bypass the issues related to the copula of $(Y,\Delta,\bX)$. In short, our intention is to discard the problem by obtaining the copula of $(Y,\bX)$ conditionally on $\Delta=1$, for which no specific technical difficulties are involved. To that end, using the notations $H^{u}$ (resp. $h^{u}$) as a shorthand for a given distribution (resp. density) conditionally on $\Delta=1$, first note that 
\begin{eqnarray}\label{eq:cond expect dev}
\E \left(\rho_{\tau}(Y-a) \frac{\Delta}{1-G_C(Y-|\bx)} \Big| \bX = \bx\right) = \int_{\RR^{+}} \rho_{\tau}(y-a) \, \frac{1}{1-G_C(y-|\bx)} \, \mathrm dF_{Y,\Delta| \bX}(y,1|\bx),
\end{eqnarray} 
 where $F_{Y,\Delta|\bX}(y,1|\bx)=\PP(Y \leq y, \Delta = 1 | \bX = \bx) = p(\bx) \, \int_{-\infty}^{y} f_{Y\bX}^{u} (z, \bx)/f^{u}(\bx) \, \mathrm dz$, with $p(\bx)=\PP(\Delta = 1 | \bX=\bx)$ and where $f^{u}_{Y\bX}$ and $f^{u}$ denote the conditional densities of $(Y,\bX)$ and $\bX$ given $\Delta=1$, respectively. Hence, using the definition of a copula function in a similar spirit as \citeauthor{NEGB13}, one obtains 
\begin{eqnarray*}
\mathrm dF_{Y,\Delta|\bX}(y,1|\bx) =  \displaystyle p(\bx) \, \frac{c^{u}_{Y\bX}\big(F^{u}_Y(y), \bF^{u}(\bx)\big)}{c^{u}_{\bX}\big(\bF^{u}(\bx)\big)} \, f^{u}_Y(y) \, \mathrm dy, 
\end{eqnarray*}
where $c^{u}_{Y\bX}(u_0, \bu) = \partial^{d+1}C_{Y\bX}^{u}(u_0,u_1,\ldots,u_d)/\partial u_0 \partial u_1 \ldots \partial u_d$ is the copula density corresponding to the copula $C^{u}_{Y\bX}$ of $(Y,\bX|\Delta=1)$, $c_{\bX}^{u}(\bu) = \partial^{d}C_{\bX}^{u}(u_1,\ldots,u_d)/\partial u_1 \ldots \partial u_d$ is the copula density of $(\bX|\Delta=1)$, and $\bF^{u}(\bx)=(F^{u}_1(x_1),\ldots,F^{u}_d(x_d))^{\mathsf{T}}$. Inserting this last expression in \eqref{eq:cond expect dev}, 
we may write 
\begin{eqnarray*}
\E \left[\rho_{\tau}(T-a)|\bX=\bx\right] = \E \left[\rho_{\tau}(Y-a) \, \frac{\Delta}{1-G_C(Y-|\bx)} \, \frac{p(\bx)}{\PP(\Delta=1)} \, \frac{c^{u}_{Y\bX}\big(F^{u}_Y(Y), \bF^{u}(\bx)\big)}{c^{u}_{\bX}\big(\bF^{u}(\bx)\big)} \right]. 
\end{eqnarray*}
Applying this equality in the context of quantile regression, interestingly, one eventually retrieves an expression analogous to \eqref{eq:CQR}: 
\begin{eqnarray}\label{eq:Cop CQR}
m_{\tau}(\bm x) = \arg\min_a  \E \left[\rho_{\tau}(Y-a) \, W(\bx) \, c^{u}_{Y\bX}\big(F^{u}_Y(Y), \bF^{u}(\bx)\big) \right],
\end{eqnarray} 
where $W(\bx)\equiv\Delta/(1-G_C(Y-|\bx))$. Note that the copula in question in \eqref{eq:Cop CQR} is determined by strictly fully observed data. Hence, standard literature on copulas can be manipulated without any censoring related constraints. Given estimators $\widehat{G}_C(\cdot|\bx)$, $\widehat{F}_Y^{u}$ and $\widehat{F}_j^{u}$ of $G_C(\cdot|\bx)$, $F_Y^{u}$ and $F_j^{u}$, $j=1,\ldots,d$, satisfying certain high-level conditions which will be given in Section \ref{section:properties}, this suggests to estimate the quantile regression in the presence of censoring by the empirical analogue of \eqref{eq:Cop CQR}, that is 
\begin{eqnarray}\label{eq:Cop CQR esti}
\widehat{m}_{\tau}(\bm x) = \arg\min_a  \sum_{i=1}^n \left[\rho_{\tau}(Y_i-a) \, \widehat{W}_i(\bx) \, \widehat{c}^{u}_{Y\bX}\big(\widehat{F}^{u}_Y(Y_i), \widehat{\bF}^{u}(\bx)\big) \right],
\end{eqnarray} 
where $\widehat{W}_i(\bx) = \Delta_i/(1-\widehat{G}_C(Y_i-|\bx))$, and where $\widehat{c}^{u}_{Y\bX}$ denotes an estimator of $c^{u}_{Y\bX}$ based on the four-step procedure described in Section \ref{sect:background}. Explicitly, 
\begin{eqnarray*}
\widehat{c}^{u}_{Y\bX}\big(\widehat{F}^{u}_Y(y), \widehat{\bF}^{u}(\bx)\big) &=& \widehat{c}^{u}_{YX_1}\big(\widehat{F}^{u}_Y(y),\widehat{F}^{u}_1(x_1)\big) \times \ldots \times \widehat{c}^{u}_{YX_d}\big(\widehat{F}^{u}_Y(y),\widehat{F}^{u}_d(x_d)\big)\\
&&\times \widehat{c}^{u}_{X_1\ldots X_d|Y}\big(\widehat{F}^{u}_{1|Y}(x_1|y),\ldots,\widehat{F}^{u}_{d|Y}(x_d|y)\big), 
\end{eqnarray*}
where any two-dimensional kernel density estimator may be used for each bivariate copula density $\widehat{c}^{u}_{YX_j}, j=1,\ldots,d$, such that condition \ref{cond:cop_hat} of Section \ref{section:properties} holds, and where $\widehat{c}^{u}_{X_1\ldots X_d|Y}$ is estimated by standard parametric vine procedures. 

The resulting quantile regression estimator in \eqref{eq:Cop CQR esti} may then be viewed as a simple weighted quantile of the observed response variable, and is therefore easy to implement in practice using the efficient quantile regression code developed by \citet{KP97} and \citet{K05}. Nonetheless, in the context of multivariate covariates, the estimation of $G_C(\cdot|\bx)$ requires further assumptions to overcome dimension related issues. Popular choices in the literature include, among others, independence between $C$ and $\bX$, the Cox model or the single-index model on $C|\bX=\bx$. Illustrations of such assumptions are treated in our simulation study.   

As an interesting property, and similarly to the case without censoring, we note that the obtained regression function estimator is automatically monotonic accross quantile levels. Applying analogous arguments to the ones adopted in the proof of Theorem 2.5 of \citet{K05}, one can indeed determine that
\begin{eqnarray}\label{eq:Cop CQR prop mono}
(\tau_2 - \tau_1)(\widehat{m}_{\tau_2}(\bx)-\widehat{m}_{\tau_1}(\bx)) \sum_{i=1}^n \widehat{W}_i(\bx) \, \widehat{c}^{u}_{Y\bX}\big(\widehat{F}^{u}_Y(Y_i), \widehat{\bF}^{u}(\bx)\big) \geq 0.
\end{eqnarray} 
Given that $\widehat{W}_i(\bx) \, \widehat{c}^{u}_{Y\bX}\big(\widehat{F}^{u}_Y(Y_i), \widehat{\bF}^{u}(\bx)\big) \geq 0$ for all $i=1,\ldots,n$, this signifies that $\widehat{m}_{\tau_2}(\bx)\geq\widehat{m}_{\tau_1}(\bx)$ for $\tau_2 \geq \tau_1$.
   
Conclusively, in parallel to what has been stated for the uncensored case, the resulting estimator $\widehat{m}_\tau(\bx)$ defines a rich class of estimators built on the many different existing methods available in the literature for estimating copula densities and marginal distributions of both complete and censored data.     

\section{Asymptotic Properties}\label{section:properties}
We establish in this section the asymptotic normality of the proposed estimator $\widehat{m}_\tau(\bx)$. To that end, we first report the set of regularity conditions as well as the required high-level conditions on all estimators involved in the expression of $\widehat{m}_\tau(\bx)$. We then develop an asymptotic representation of our estimator for a general $d$-variate covariate. As the latter will result in a somewhat unpleasant expression for the asymptotic bias and variance for a general multivariate covariate, and given that the analytical reasoning is similar in spirit, we eventually restrict ourselves to the detailed asymptotic expression for the case $d=2$.

For a fixed but arbitrary point of interest $\bx$ in the support of $\bX$, denoted by supp($\bX$), let us suppose that there exists a neighborhood $\mathcal{V}_{\bF^{u}(\bx)}$ of $\bF^{u}(\bx)$ such that the following regularity conditions hold:   
	\begin{itemize}
	\item[\namedlabel{cond:true f}{(C1)}] The conditional distribution $F_{T|\bX}$ of $T$ given $\bX$ admits a conditional density $f_{T|\bX}$ that is continuous, strictly positive and bounded uniformly on $\RR \times \text{supp}(\bX)$.
	\item[\namedlabel{cond:pt x}{(C2)}] The point of interest $\bx$ is such that $\bF^{u}(\bx) \in (0,1)^d$ and $\PP(\Delta=1|\bx)>0$. Furthermore, $0<c_\bX^{u}(\bF^{u}(\bx))<\infty$, $\sup_{t\in \RR} c_{Y\bX}^{u}(F_Y^{u}(t),\bF^{u}(\bx))<\infty$ and $\inf_{t\in \RR} c_{Y\bX}^{u}(F_Y^{u}(t),\bF^{u}(\bx))>0$.
	\item[\namedlabel{cond:G_C}{(C3)}] The point $m_\tau(\bx) \in \RR$ satisfies $G_C(m_{\tau}(\bx)+\delta|\bx)<1$, for some $\delta >0$.
	
	\item[\namedlabel{cond:techn}{(C4)}] Denote $\epsilon \equiv \epsilon(\bx,\tau)=Y-m_{\tau}(\bx)$ and define $\psi_{\tau}(u)=\tau-I(u \le 0)$. Then,
																			\begin{itemize}
																				\item[(i)]  $\E(|\psi_{\tau}(\epsilon)|\,W(\bx))<\infty$. 
																				\item[(ii)] $\E\left[\psi_{\tau}(\epsilon)\,W(\bx)\,c^{u}_{Y\bX}\big(F^{u}_Y(Y), \bF^{u}(\bx)\big)\right]^2 < \infty$.
																			\end{itemize}
	
\end{itemize}
Concerning the high-level conditions, it is assumed that the multivariate copula density $c^{u}_{Y\bX}$ is estimated using the proposed four-step strategy of Section \ref{sect:background}, and that, for simplicity, the $d$ bivariate kernel copula estimators of step \ref{cop:step2} are based on the same bandwidth $\bm{H}=h^2\bm{I}$ for a certain $h>0$. The following conditions are then assumed to hold: 
		
\begin{itemize}
	\item[\namedlabel{cond:marg}{(C5)}] The marginal c.d.f. estimators are such that:
																			\begin{itemize}
																				\item[(i)]  $\sup_{t\in \RR}\left|\widehat{F}_Y^{u}(t)-F_Y^{u}(t)\right| = O_p(n^{-1/2})$.
																				\item[(ii)] $\widehat{\bF}^{u}(\bx)-\bF^{u}(\bx) = O_p(n^{-1/2})$, where $\widehat{\bF}^{u}(\bx) = (\widehat{F}^{u}_1(x_1),\ldots,\widehat{F}^{u}_d(x_d))^{\mathsf{T}}$ and $\widehat{F}^{u}_j$ is an estimator of $F^{u}_j$.
																			\end{itemize}
	\item[\namedlabel{cond:G_C_hat}{(C6)}] ${\displaystyle \sup_{t\leq \tau_{F_Y}}|\widehat{G}_C(t|\bx)-G_C(t|\bx)|=o_p((nh^2)^{-1/2})}$, and $\tau_{F_Y}< \tau_{G_C}$, where $\tau_{F_Y} = \sup\{t : F_Y(t) < 1\}$ and $\tau_{G_C} = \sup\{t : G_C(t) < 1\}$. 
	\item[\namedlabel{cond:cop_hat}{(C7)}] The multivariate copula estimator is such that: 
																		\begin{itemize}
																				\item[(i)] $\sup_{\,t\,\in\,\RR} \left|\widehat{c}^{u}_{YX_j}\big(F^{u}_Y(t), F_j^{u}(x_j)\big)-c^{u}_{YX_j}\big(F^{u}_Y(t), F_j^{u}(x_j)\big)\right|=o_p(1)$, $j=1,\ldots,d$, where $x_j$ is the $j$-th coordinate of $\bx$.
																				\item[(ii)] $\sup_{u_0\,\in\,(0,1)}\sup_{\,\bm{u}\,\in\, \mathcal{V}_{\bF^{u}(\bx)}}\left|\partial_j \, \widehat{c}^{u}_{Y\bX}(u_0, \bm{u})\right| = O_p(1)$, $j=1,\ldots,d+1$, where $\partial_j$ denotes the partial derivative with respect to the $j$-th argument.
																			\end{itemize}	
\end{itemize}

Assumption \ref{cond:true f} is standard in the context of quantile regression estimation. As for condition \ref{cond:pt x}, this is similar to assumption (C3)-(i) in \citet{NEGVK15} for the simplified case with no censoring, with an additional requirement on the conditional censoring probability that is resulting from the initial transformation of synthetic observations. Assumption \ref{cond:G_C} is likewise emanating from the handling of censoring through these observations, and is rather usual in survival analysis. Note that, in the quantile regression framework, the latter assumption amounts to defining a natural upper bound for the quantile of interest that can be studied. Assumption \ref{cond:techn} reports a set of technical conditions to be met.

As regards conditions \ref{cond:marg}-\ref{cond:cop_hat}, \ref{cond:marg} is routinely made in the copula framework. For instance, it is readily satisfied for the empirical distributions when only uncensored observations are taken into account, and their rescaled versions which are prominent in the copula literature. Assumption \ref{cond:G_C_hat} imposes restrictions on the estimator one may consider for the conditional distribution of the censoring variable and is, for instance, fulfilled for a simple Kaplan-Meier estimator for $G_C$ (see e.g. Theorem 2.1 in \citet{CL97} for sufficient and necessary conditions for \ref{cond:G_C_hat}). Lastly, the uniform consistency of the kernel density estimator required by assumption \ref{cond:cop_hat} is, for instance, alluded to in \citet{GCP14} for the probit-transformed copula estimator.

We now state the main result of this section that holds for a general $d$-dimensional covariate vector and for all bivariate kernel copula estimators based on the same bandwidth $h$. In practice, however, it may be recommended to adopt an unconstrained and non-diagonal bandwidth matrix, as is detailed in Section 4 of \citeauthor{GCP14}. Nevertheless, when considering this general situation, the theoretical results become less tractable while equivalent in nature to the simplified situation considered here.  

\begin{theorem}\label{theorem1}
Let $h \equiv h_n \rightarrow 0$ be the common bandwidth of the $d$ bivariate kernel copula density estimators. For $h$ satisfying $nh^2 \rightarrow \infty$ as $n \rightarrow \infty$, and under assumptions \ref{cond:true f}-\ref{cond:cop_hat}, we have 
\begin{multline*}
\left(nh^2\right)^{1/2}(\hat{m}_{\tau}(\bx)-m_{\tau}(\bx))=\\ \frac{w(\bx)}{f_{T|\bX}(m_{\tau}(\bx)|\bx)} \, \frac{\left(nh^2\right)^{1/2}}{n}\sum_{i=1}^n\psi_{\tau}(\epsilon_i)W_i(\bx)\big[\widehat{c}^{u}_{Y\bX}\big(F^{u}_Y(Y_i), \bF^{u}(\bx)\big) - c^{u}_{Y\bX}\big(F^{u}_Y(Y_i), \bF^{u}(\bx)\big) \big] + o_p(1),
\end{multline*}
where $f_{T|\bX}$ is the conditional density of $T$ given $\bX$ and $w(\bx) = p(\bx)/\big[\PP(\Delta=1)c^u_{\bX}(\bF^u(\bx))\big]$.
\end{theorem}
Theorem \ref{theorem1} implies, quite naturally, that the asymptotic behaviour of $\hat{m}_{\tau}(\bx)$ will be characterized by the properties of the copula estimator, specifically through its nonparametric feature, provided that the estimation of $\widehat{G}_C(\cdot|\bx)$ is `reasonable' when confronted to a multidimensional covariate vector (assumption \ref{cond:G_C_hat}). In particular, this suggests that the detailed discussion of \citeauthor{GCP14} about the asymptotic bias and variance of their distinctive bivariate copula estimators may be transcribed in our context. 

Additionally, Theorem \ref{theorem1} also covers an asymptotic representation of the copula-based quantile regression estimator when all responses are fully observed. In this situation, one would indeed obtain a similar result for the proposed semiparametric procedure, with the removal of all censoring related terms, that is $w(\bx)$, $W_i(\bx)$, $i=1,\ldots,n,$ and the superfluous conditioning on $\Delta=1$ for the copula densities and marginal distributions. 

We now consider a detailed asymptotic representation of our estimator for the simplified case where $d=2$, and for a general nonparametric estimator of the bivariate copula densities. For convenience, we use the notation $c^{u}_k$ as a shorthand for $c^{u}_{YX_k}$, and similarly for other functions depending on $(Y,\Delta,X_k)$, $k=1,2$.
\begin{corr}\label{corr1}
Suppose that the assumptions of Theorem \ref{theorem1} hold for the case $d=2$. Furthermore, suppose that the bivariate nonparametric copula estimators of $c^{u}_{1}$ and $c^{u}_{2}$ are such that 
\begin{multline*}
\left(nh^2\right)^{1/2}\Big(\widehat{c}^{u}_{k}(u_0,u_k)-c^{u}_{k}(u_0,u_k)-h^2b_{k}(u_0,u_k)\Big)=\frac{1}{\sqrt{n}}\sum_{j=1}^nZ^{nj}_{k}(u_0,u_k)+o_p(1), \\ \forall u_k \in  (0,1), \text{ uniformly in } u_0 \in  (0,1), \text{ for } k=1,2,
\end{multline*}
for some some deterministic function $b_{k}(u_0,u_k)$, and for some function $Z^{nj}_{k}(u_0,u_k)$ depending on $(Y_j, \Delta_j, X_{kj})$ and possibly on $n$, satisfying $\E\left(Z^{nj}_{k}(u_0,u_k)\right)=0$, for all $u_0, u_k \in (0,1)$. 

Define 
\begin{align*}
\widetilde{Z}^{ni}(u_0,\bm{u}) &= \left[Z^{nj}_1(u_0,u_1)  c^{u}_{2}\big(u_0, u_2\big) + Z^{nj}_2(u_0,u_2) c^{u}_{1}\big(u_0, u_1\big) \right] c^{u}_{X_1X_2|Y}(u_1,u_2|u_0),\\ 
b_{Y\bX}(u_0,\bm{u}) &= \left[ b_1(u_0,u_1) c^{u}_{2}\big(u_0, u_2\big) + b_2(u_0,u_2) c^{u}_{1}\big(u_0, u_1\big)  \right] c^{u}_{X_1X_2|Y}(u_1,u_2|u_0),\\
\lambda_n\left(Y_i,\Delta_i,\bX_i,\bx\right) &=\E\left[\psi_{\tau}(\epsilon)W(\bx)\widetilde{Z}^{ni}\left(F^{u}(Y),\bF^{u}(\bx)\right)|Y_i,\Delta_i,\bX_i\right], i=1,\ldots,n.
\end{align*} 

Suppose furthermore that the following technical conditions hold:
\begin{itemize}
	\item[\namedlabel{cond:techn_biais}{(C8)}] $\E \left[\psi_{\tau}(\epsilon)W(\bx) b_{Y\bX}\left(F^{u}(Y),\bF^{u}(\bx)\right)\right]^2 < \infty$. 
	\item[\namedlabel{cond:techn_Vstat}{(C9)}] $\E\left[\psi_{\tau}(\epsilon_i)W_i(\bx)\widetilde{Z}^{nj}\left(F^{u}(Y_i),\bF^{u}(\bx)\right)\right]^2 = o(n)$ for all $i,j=1,\ldots,n$, where the expectation is taken with respect to $(Y_i,\Delta_i)$ and $(Y_j,\Delta_j,\bX_j)$.
\end{itemize} 

Then, the copula-based quantile regression estimator at any point of interest $\bx$ satisfying \ref{cond:true f}-\ref{cond:techn_Vstat} is such that 
\begin{eqnarray*}
\left(nh^2\right)^{1/2}\Big(\widehat{m}_{\tau}(\bx)-m_{\tau}(\bx)-h^2B(\bx)\Big) \stackrel{\mathcal{L}}{\longrightarrow} \mathcal{N}\big(0,\sigma^2(\bx)\big),
\end{eqnarray*}
where 
\begin{align*}
B(\bx) &= \frac{w(\bx)}{f_{T|\bX}(m_{\tau}(\bx)|\bx)} \E \Big(\psi_{\tau}(\epsilon)W(\bx)b_{Y\bX}\big(F^{u}(Y),\bF^{u}(\bx)\big)\Big)\\
\text{and}\;\sigma^2(\bx) &= \frac{w^2(\bx)}{f_{T|\bX}^2(m_{\tau}(\bx)|\bx)} \lim_{n \rightarrow \infty}\E\,(\lambda_n(Y,\Delta,\bX,\bx)^2).
\end{align*}
\end{corr}
Corollary \ref{corr1} reports the asymptotic normality of our estimator at the expected convergence rate, implied by the nonparametric estimation of the bivariate copula densities. Depending on the choice in step \ref{cop:step2} of the kernel density estimator fulfilling the conditions of Corollary \ref{corr1}, simple plug-in of the expression of $Z_k^{nj}$, $k=1,2$, in all quantities built upon the latter may then lead to the detailed, although arduous, expressions of the asymptotic bias and variance of the proposed estimator. 

Furthermore, as this had yet to be covered, it is worth stressing out that Corollary \ref{corr1} also encompasses the asymptotic normality of the suggested estimator based on semiparametric vine copulas with strictly complete data. Similarly to what has been stated for Theorem \ref{theorem1}, one is indeed only required to withdraw all censoring related elements from Corollary \ref{corr1} to obtain the expressions of the asymptotic bias and variance of the proposed semiparametric quantile regression estimator for complete observations.
 

\section{Simulation Study}\label{section:simulations}
In this section, we assess the practical finite-sample performance of the proposed methodology by means of Monte Carlo simulations. For this purpose, we first present a brief numerical study to further motivate the semiparametric copula strategy we intend to adopt for multivariate problems for both complete and censored observations. Secondly, focussing on survival data, we illustrate the flexibility of our estimator based on the proposed copula modelling by showing promising results with respect to competitors, including when the generated scenario is to the advantage of the latter. All the simulations are carried out using the statistical computing environment R (\citet{R14}) and its freely accessible packages.

\subsection{Assessing the semiparametric copula estimation}\label{subsection:copula strategy}
This first section aims at numerically illustrating the choice of our semiparametric copula estimation strategy. For this purpose, we consider two distinctive data generating processes (DGP) and compare our methodology with fully parametric and nonparametric procedures one might consider for the estimation of a multivariate copula density. For the general simulation settings, we consider $B=500$ repetitions of each DGP; three (average) levels of censoring (0\%, 30\% and 50\%), three sample sizes ($n \in \{100,200, 400\}$) and the quantile level of interest $\tau = 0.3$. As the object of interest here is the copula modelling, when censoring is introduced, we only consider the simple case of independence between the censoring variable and the covariate vector in order to keep the estimation of $\widehat{G}_C$ needed for \eqref{eq:Cop CQR esti} uncomplicated, that is, using the Kaplan-Meier estimator. The detailed DGPs are as follows:
\begin{enumerate}
	\item[$\bullet$] \textbf{DGP A}: $(F_T(T),F_{1}(X_1),F_{2}(X_2)) \sim$ Gaussian copula with parameters $(\rho_{T1},\rho_{T2},\rho_{12}) = (0.3,0.9,0.5)$. Given standard uniform marginal distributions for all three variables, the resulting true quantile regression may be calculated as $m_\tau(\bx) = \Phi\big(-0.2\Phi^{-1}(x_1) + \Phi^{-1}(x_2) + 0.4\Phi^{-1}(\tau)\big)$ (see \citet{NEGVK15}). To include censoring, we introduce the variable $C \sim U[0,M]$, where the parameter $M$ is computed in order to obtain the desired average censoring proportion ($M=5/3$ for 30\% and $M=1$ for 50\%).
	\item[$\bullet$] \textbf{DGP B}: $(F_T(T),F_{1}(X_1),F_{2}(X_2),F_{3}(X_3)) \sim$ Gaussian copula with parameters $(\rho_{T1},\rho_{T2},\rho_{T3}$, $\rho_{12},\rho_{13},\rho_{23}) = (0.3,0.9,0.7,0.5,0.25,0.5)$. The resulting true quantile regression for standard uniform marginal distributions is determined as $m_\tau(\bx) = \Phi\big(-0.2\Phi^{-1}(x_1) + 0.83\Phi^{-1}(x_2) + 0.33\Phi^{-1}(x_2) + 0.27\Phi^{-1}(\tau)\big)$. The censoring variable is $C \sim U[0,M]$ ($M=5/3$ for 30\% and $M=1$ for 50\% censoring).
\end{enumerate} 
For any general copula-based regression estimator, the marginal distribution estimations are performed, as suggested in Section \ref{sect:background}, using rescaled versions of the empirical distributions:
\begin{align*}
\widehat{F}^{u}_Y(y) = \frac{1}{n^{u}+1}\sum_{i=1}^{n} \Delta_i \ind(Y_i \leq y) && \widehat{F}^{u}_{j}(x_j) = \frac{1}{n^{u}+1}\sum_{i=1}^{n} \Delta_i \ind(X_{ij} \leq x_j), j=1,\ldots,d,
\end{align*}
where $n^{u} = \sum_{i=1}^n \Delta_i$ is the number of uncensored observations. 

For the distinctive copula-based estimators, we consider the following procedures:
\begin{enumerate}
	\item[$\widehat{m}_{cop,\, \tau}^{SP}$]: semiparametric estimation strategy detailed in Section \ref{sect:background}. That is, we first estimate the $d$ bivariate copulas of interest employing the probit transformation technique of \citet{GCP14} coupled with local likelihood estimation based on quadratic polynomials. To that end, we follow their proposed nearest-neighbor bandwidth selection procedure. Concerning the estimation of the $d$-dimensional `noisy' copula density \eqref{eq:multivariate cop part 2}, as mentioned above, we apply standard vine techniques using the R package \texttt{VineCopula}. Specifically, we adopt one automatically selected tree structure for the simplified decompositon of the copula density among many R-vine candidate structures (see \citet{DBCK13}), and subsequently determine the appropriate pair-copula family to be selected and parametrically estimated. The selection criterion for bivariate copulas is chosen to be the Akaike information criterion (AIC), which revealed to be adequate in the R-vine context (see \citet{B10}, chap. 5), and ten potential family candidates, together with their rotations, are considered: eight of them are Archimedian (Clayton, Gumbel, Frank, Joe, Clayton-Gumbel, Joe-Gumbel, Joe-Clayton and Joe-Frank), and the last two are elliptical (Gaussian and Student $t$).
	\item[$\widehat{m}_{cop,\, \tau}^{NP}$]: fully nonparametric estimation of the $d$-dimensional copula using vine techniques. Specifically, while the vine structure is kept identical, here all bivariate building blocks are estimated using the local likelihood technique based on probit-projected data with, here again, the bandwidth selection procedure of \citeauthor{GCP14} (as is studied in \citet{NC15}). Given its fully nonparametric nature, it should be mentioned that this estimator is not covered by the theoretical results of Section \ref{section:properties}.
	\item[$\widehat{m}_{cop,\, \tau}^{P}$]: fully parametric estimation of the $d$-dimensional copula density, where all bivariate copulas are estimated using the previously mentioned candidate families and selection criteria. However, unlike the above-mentioned estimators, we do not force here any structure for the vine decomposition. As a consequence, no explicit distinction is imposed between dependence of interest and noisy dependence. Instead, one data-driven selected structure is adopted, regardless of the arguments of Section \ref{sect:background}. This will allow us to analyse the impact of such dependence distinction in our regression context, as is discussed below. Finally, as it is the case for $\widehat{m}_{cop,\, \tau}^{NP}$, this estimator is not covered by the asymptotic theory of Section \ref{section:properties}.
\end{enumerate}

Both DGPs concentrate on the situation where the dependence structure between the response variable and the covariate vector is characterized by a parametric copula. In such circumstances, $\widehat{m}_{cop,\, \tau}^{P}$ will have a critical advantage, and may serve in order to evaluate the impact of the nonparametric part of the estimation scheme, especially when the dimension of the covariate vector increases. As a performance criterion, we consider here the empirical integrated mean squared error (IMSE), defined as 
$$IMSE(\widehat{m}_\tau(\bx)) = \frac{1}{N} \sum_{i=1}^{N} \left( \frac{1}{B} \sum_{b=1}^B \big(\widehat{m}_\tau^{(b)}(\bx_i) - m_\tau(\bx_i)\big)^2\right),$$ 
where $\{\bx_i, i=1,\ldots,N\}$ is a generated random sample of size $N = 10$ serving as an evaluation set spread on the domain of $\bX$, and $\widehat{m}_\tau^{(b)}(\cdot)$ denotes the regression estimation for the $b$-th simulated sample.

The results of our simulation study are summarized in Table \ref{table:Copula Strategy}. Based on these, we detail our analysis in two parts, as the outcomes of our study offer relevant information on both the copula decomposition choice and the type of bivariate estimators one may adopt in the multivariate setting. Note that, for both DGPs, as the dependence structure is specified by a Gaussian copula, the simplifying assumption intrinsic to the vine decomposition is here applicable (see Theorem 4 in \citet{SJC13}). In our context, this means that any observed difference between copula strategies is not to be attributed to a possible violation of the underlying simplifying assumption. 

\begin{table}[t!]
\centering
\begin{tabular}{ccc||c|c|c}
DGP & $n$ & $p_c$ & $\widehat{m}_{cop,\, \tau}^{P}$ & $\widehat{m}_{cop,\, \tau}^{SP}$ & $\widehat{m}_{cop,\, \tau}^{NP}$\\
\hline
\hline
\multirow{9}{*}{\textbf{A}} & \multirow{3}{*}{100}   & 0   & 1.442 & 1.486 & 2.369 \\
                            &                        & 0.3 & 2.203 & 2.173 & 3.167 \\
                            &                        & 0.5 & 3.537 & 3.548 & 4.135 \\ \cline{2-6}
                            & \multirow{3}{*}{200}   & 0   & 0.689 & 0.737 & 1.462 \\
                            &                        & 0.3 & 0.990 & 1.016 & 1.905 \\
                            &                        & 0.5 & 1.887 & 1.664 & 2.581 \\ \cline{2-6}
														& \multirow{3}{*}{400}   & 0   & 0.337 & 0.371 & 0.987 \\
                            &                        & 0.3 & 0.503 & 0.525 & 1.243 \\
                            &                        & 0.5 & 0.915 & 0.863 & 1.600 \\ 
\hline
\hline
\end{tabular}
\quad
\begin{tabular}{ccc||c|c|c}
DGP & $n$ & $p_c$ & $\widehat{m}_{cop,\, \tau}^{P}$ & $\widehat{m}_{cop,\, \tau}^{SP}$ & $\widehat{m}_{cop,\, \tau}^{NP}$\\
\hline
\hline
\multirow{9}{*}{\textbf{B}} & \multirow{3}{*}{100}   & 0   & 1.192 & 1.416 & 3.307 \\
                            &                        & 0.3 & 2.036 & 2.459 & 3.972 \\
                            &                        & 0.5 & 5.055 & 6.552 & 7.038 \\ \cline{2-6}
														& \multirow{3}{*}{200}   & 0   & 0.560 & 0.658 & 2.609 \\
                            &                        & 0.3 & 0.850 & 0.945 & 2.857 \\
                            &                        & 0.5 & 1.686 & 2.108 & 3.647 \\ \cline{2-6}
                            & \multirow{3}{*}{400}   & 0   & 0.259 & 0.354 & 2.140 \\
                            &                        & 0.3 & 0.403 & 0.487 & 2.266 \\
                            &                        & 0.5 & 0.738 & 0.916 & 2.497 \\ 																			
\hline
\hline
\end{tabular}
\caption{Simulation results expressed in terms of $IMSE \times 1000$ for the estimation of $m_\tau(\bx)$ in DGP \textbf{A} and \textbf{B}. The number of repetitions operated is $B=500$ for sample sizes $n \in \{100,200,400\}$, average censoring proportions $p_c \in \{0,0.3,0.5\}$ and quantile level $\tau=0.3$.}
\label{table:Copula Strategy}
\end{table}

Focussing first on our decomposition strategy, we note that, as expected, $\widehat{m}_{cop,\, \tau}^{P}$ globally outperforms $\widehat{m}_{cop,\, \tau}^{SP}$ and $\widehat{m}_{cop,\, \tau}^{NP}$. However, strikingly enough, this is not observed for DGP \textbf{A} for different censoring proportions, where $\widehat{m}_{cop,\, \tau}^{SP}$ details better results. This is interpreted here as evidence for the validity of our arguments regarding the decomposition choice: as the censoring proportion grows, the number of observations actually entering the copula estimation becomes more moderate, hereby implying two opposite effects in this context. First, the propagation of estimation approximations tends to be more important, signifying that the further we decompose, the more sensitive becomes the estimation of the involved bivariate copulas as these are tributary of the quality of previously estimated bivariate blocks. Using a purely data-driven decomposition may then result in a poor fit of the (conditional) copula of the response variable with one of the covariates, as it is not required that the latter would be primarily treated. This is interpreted as the reason why $\widehat{m}_{cop,\, \tau}^{SP}$ is able to outperform $\widehat{m}_{cop,\, \tau}^{P}$, admittedly by a small amount, when censoring increases for a fixed sample size $n \in \{200,400\}$, even though the simulated scenario is issued from a purely parametric copula. However, on the other hand, when observations are more scarse, it is well-known that nonparametric estimations become more sensitive than parametric counterparts. This explains why the estimation results for $\widehat{m}_{cop,\, \tau}^{SP}$ are not superior to those of $\widehat{m}_{cop,\, \tau}^{P}$ for $n=100$ with 50\% censoring, as the former requires the nonparametric estimation of two bivariate copulas, whose complexity compared to $\widehat{m}_{cop,\, \tau}^{P}$ seems to override the positive effects of our decomposition choice. Overall, these noteworthy results for DGP \textbf{A} illustrate the effectiveness of our proposed copula decomposition in the regression context. When augmenting the covariate vector dimension, the price of estimating now three nonparametric bivariate copulas quite logically exceeds the potential gain of concentrating efforts on the dependence of interest. This is identified in DGP \textbf{B}. 

Concentrating now on the modelling choice for the noisy dependence, the comparison between $\widehat{m}_{cop,\, \tau}^{SP}$ and $\widehat{m}_{cop,\, \tau}^{NP}$ offers valuable information, particularly in DGP \textbf{A}, as the only distinctness here is the estimation of a unique bivariate copula density $c^{u}_{X_1X_2|Y}$. Visibly, the implication of keeping a nonparametric approach for this part seems to be rather severe. This finding clearly also applies when the dimension of the covariate grows.

Conclusively, the recommended copula modelling strategy seems to propose an adequate trade-off between preventing the serious effects of a purely nonparametric estimation and providing the flexibility needed to overcome possible shortcomings associated to a purely parametric alternative. 

\subsection{Comparison with other estimation methods}
The objective of this second section is to provide a comparison study between the proposed copula-based methodology and existing competitors for survival data. On a general note, given that for multidimensional covariates an appropriate estimation of the conditional distribution $G_C(\cdot|\bx)$ for the weights $W(\bx)$ in \eqref{eq:Cop CQR esti} may be of crucial influence, we consider distinctive scenarios contrasting in the impact on $\widehat{G}_C(\cdot|\bx)$ in order to provide a sufficiently broad view on the performance of our methodology.   

We examine 2 general simulation models, with $B = 500$ repetitions of each; two (average) levels of censoring (30\% and 50\%), two sample sizes ($n=200$ and $n=400$) and four values for the quantile level of interest ($\tau \in \{0.1,0.3,0.5,0.7\}$). Specifically, we consider general data arising from a Cox regression model. Based on the hazard function (defined as $h(t) = f_T(t)/(1-F_T(t))$), this prominent model for analysing survival times specifies that $h(t|\bX_i) = h_0(t) \exp(\beta^{\mathsf{T}} \bX_i), i=1,\ldots, n$, where $h_0(t)$ is the so-called baseline hazard function. In this instance, given a nonnegative time-to-event variable, simple algebraic manipulations show that the general conditional quantile regression can be written as $m_\tau(\bx) = H_0^{-1}\big(-\log(1-\tau) \, \exp(-\beta^{\mathsf{T}} \bx) \big)$, where $H_0(t) \equiv \int^t_0 h_0(s) \mathrm ds$. For every proposed setting, we choose $\beta=(1,-3/4,1/2,1/4,-3/5)^{\mathsf{T}}$ and $d=5$. Furthermore, the baseline distribution is set to be standard exponential, and, consequently, for a given vector $\bx$ the $\tau-$th conditional quantile function is given by 
$$
m_\tau(\bx) = -\log(1-\tau) \, \exp(-\beta^{\mathsf{T}} \bx). 
$$

The distinction between our simulation scenarios is to be found in the covariate vectors that are taken into account. Let us consider 5 covariates $X_j$, $j=1,\ldots,5$, with standard uniform distributions. In order to allow for covariate dependence, we simulate $(X_1,\ldots,X_5)$ form a 5-variate Gaussian copula, with parameters $(\rho_{12},\rho_{13},\ldots,\rho_{45})=(0.3, 0.4, 0.5, 0.6, 0.7, 0.3, 0.4, 0.5, 0.6, 0.7)$. To distinguish our scenarios, we consider two covariate vectors given by $\bX^{(1)} = (X_1,X_2,\ldots,X_5)$ and $\bX^{(2)} = (X_1,\exp(X_2),X_3,X_4,X_5)$, for which the resulting quantile regressions are both single-index models in $\bX^{(1)}$ and $\bX^{(2)}$, respectively. The motivation of our simulation study is to highlight the performance of our procedure when its competing estimators are both in an ideal situation and a slightly altered version of it. 

\subsubsection[Model 1]{Model 1: covariate vector $\bX^{(1)}$}

We first consider the simple case of data following a Cox regression model issued from covariate vector $\bX^{(1)}$. The distributions and parameter values of the censoring variables define the following scenarios:
\begin{enumerate}
	\item[$\bullet$] \textbf{DGP C}: $C \sim \text{Exp}(\lambda_C)$,  with $\lambda_C$ independent of $\bX^{(1)}$. To attain the desired average censoring proportions, we fix $\lambda_C=0.464$ and $\lambda_C=1.083$ (corresponding to approximately 30\% and 50\% censoring in average, respectively). The exact conditional censoring probability given $\bX^{(1)}=\bx$ is calculated as $\lambda_C/(\lambda_C+\exp(\beta^{\mathsf{T}} \bx))$ and is hence a decreasing function of $\beta^{\mathsf{T}} \bx$, making us conjecture better results for higher values of $\beta^{\mathsf{T}} \bx$. Note that in this scenario, $G_C(\cdot|\bx)$ boils down to $G_C(\cdot)$, for which the Kaplan-Meier estimator is suited. 
	\item[$\bullet$] \textbf{DGP D}: $C \sim \text{Exp}(\lambda_{C}(\bx))$, with $\lambda_{C}(\bx) = 3/7\times \exp(\beta^{\mathsf{T}} \bx)$ and $\lambda_{C}(\bx) = \exp(\beta^{\mathsf{T}} \bx)$ corresponding to 30\% and 50\% censoring, respectively. In this scenario, the conditional censoring probability is independent of $\bx$, and an adequate estimation of $G_C(\cdot|\bx)$ is fulfilled with a Cox model hypothesis for the relationship between the covariates and the censoring variable itself.
\end{enumerate}

For comparison purposes, we consider the four following estimators:
\begin{enumerate}
	\item[$\widehat{m}_{cox,\, \tau}$]: parametric estimator exploiting the information related to the parametric Cox model setting. This estimator will serve as a reference for the ideal, yet unknown in practice, situation. Specifically, $\widehat{m}_{cox,\, \tau}(\bx) = -\log(1-\tau)\exp(-\widehat{\beta}^{\mathsf{T}} \bx)/\widehat{\lambda}_{T_0}$, where $\widehat{\beta}$ is estimated by maximum partial likelihood, and $\widehat{\lambda}_{T_0}$ is the maximum likelihood estimator of the exponential baseline distribution. 
	\item[$\widehat{m}_{si,\, \tau}$]: Single-index regression estimator studied in \citet{BEGVK14}, where the censoring distribution is supposed to be independent of $\bx$. For the univariate nonparametric part of the estimation process, 10 different bandwidths are selected ($h \in \{0.05,0.1,\ldots,0.5\}$), and the optimal choice is performed using the described leave-one-out cross-validation procedure. Note that the quantile regression of interest $m_\tau(\bx)$ here is indeed a single-index model in $(X_1,X_2,\ldots,X_5)$. This should provide a critical advantage to the performance of $\widehat{m}_{si,\, \tau}$. 
	\item[$\widehat{m}_{cop,\, \tau}^{(\independent)}$]: our copula-based estimator from estimation equation \eqref{eq:Cop CQR esti}, where the conditional distribution $G_C(\cdot|\bx)$ is supposed to be independent of $\bx$ and thereby estimated by the classical (unconditional) Kaplan-Meier estimator.
	\item[$\widehat{m}_{cop,\, \tau}^{(cox)}$]: our copula-based estimator, where the relationship between the covariates and the censoring variable is supposed to follow a Cox regression model. Namely, $\widehat{G}_C(c|\bx)$ is estimated by $1-\exp\left(-c \exp(\widehat{\beta}^{\mathsf{T}}_C\, \bx) /\widehat{\lambda}_{C_0} \right)$.
\end{enumerate}

Following the arguments of Section \ref{sect:background} and the results of Section \ref{subsection:copula strategy}, both copula-based procedures are implemented employing a semiparametric estimation for the $d$-variate copula built on the aforementioned candidate families and selection criterions.

In order to compare the studied estimators' performance, we consider an integrated version of the median absolute estimation error, that is 
\begin{eqnarray*}
	IMAE(\widehat{m}_\tau(\bx)) = \frac{1}{N} \sum_{i=1}^N \text{med}^{(B)} \big(|\widehat{m}_\tau(\bx_i) - m_\tau(\bx_i)|\big),
\end{eqnarray*}  
where $\widehat{m}_\tau$ is a generic estimator of $m_\tau$, $\{\bx_i, i=1,\ldots,N\}$ is an evaluation set corresponding to a generated random sample of size $N = 10$, spread on the domain of $\bX^{(1)}$, and $\text{med}^{(B)}$ denotes the median taken over all $B=500$ simulations. The choice for this robust $L_1$-type of measure is motivated by the fact that the optimization routines involved in the single-index procedure may yield very unlikely results with a small probability, hereby strongly disadvantaging the estimator when considering a $L_2$-type of error measure. The same reasoning is underlying the determination of a robust dispersion measure on estimation errors, taken as the averaged interquartile range. More precisely, we choose $N^{-1} \sum_{i=1}^N \left(Q_3^{(B)} \big(|\widehat{m}_\tau(\bx_i) - m_\tau(\bx_i)|\big) - Q_1^{(B)} \big(|\widehat{m}_\tau(\bx_i) - m_\tau(\bx_i)|\big)\right)$, where $Q_3^{(B)}$ and $Q_1^{(B)}$ stand, respectively, for the third and first quartiles taken over all simulations.  

The results of our simulation study for this model are reported in Tables \ref{table:model 1 censoring 30} and \ref{table:model 1 censoring 50} for 30\% and 50\% of censoring, respectively. As expected, the Cox regression estimator $\widehat{m}_{cox,\, \tau}$, serving as a reference case here, outperforms the other estimators for both scenarios and every sample size, quantile level and censoring percentage. More interestingly, while the single-index estimator $\widehat{m}_{si,\, \tau}$ quite logically displays better overall results than our copula-based estimators, it is worth noticing that the difference seems to fade away when moving to higher quantile levels, especially for higher levels of censoring percentage. This is considered as a first encouraging result for the copula-based estimators as the simulated model is very strongly to the advantage of $\widehat{m}_{si,\, \tau}$. This indicates that our proposed procedure is flexible enough to compete with $\widehat{m}_{si,\, \tau}$ in its ideal setup when the number of observations actually entering the estimation scheme becomes moderate, that is when censoring percentage is important and the quantile level of interest is high. Of course, if this is not the case, there is no a priori reason to believe that the copula-based estimators could outperform $\widehat{m}_{si,\, \tau}$ when the latter is considered in its optimal setting.

\begin{table}[t!]
\centering
\begin{tabular}{ccc||c|c|c|c}
\multicolumn{7}{c}{Censoring = 30\%} \\
DGP & $n$ & $\tau$ & $\widehat{m}_{cox,\, \tau}$ & $\widehat{m}_{si,\, \tau}$ & $\widehat{m}_{cop,\, \tau}^{(\independent)}$ & $\widehat{m}_{cop,\, \tau}^{(cox)}$\\
\hline
\hline
\multirow{8}{*}{\textbf{C}} & \multirow{4}{*}{200} & 0.1 & 0.016 (0.019) & 0.039 (0.053) & 0.043 (0.059) & 0.043 (0.057) \\
                            &                      & 0.3 & 0.054 (0.063) & 0.083 (0.103) & 0.101 (0.124) & 0.102 (0.123) \\
                            &                      & 0.5 & 0.105 (0.123) & 0.149 (0.185) & 0.174 (0.199) & 0.175 (0.194) \\
                            &                      & 0.7 & 0.183 (0.213) & 0.280 (0.373) & 0.273 (0.299) & 0.279 (0.305) \\\cline{2-7}
                            & \multirow{4}{*}{400} & 0.1 & 0.011 (0.013) & 0.030 (0.038) & 0.033 (0.044) & 0.033 (0.043) \\
                            &                      & 0.3 & 0.036 (0.043) & 0.070 (0.078) & 0.079 (0.097) & 0.081 (0.098) \\
                            &                      & 0.5 & 0.069 (0.084) & 0.115 (0.140) & 0.130 (0.160) & 0.133 (0.159) \\
                            &                      & 0.7 & 0.120 (0.146) & 0.221 (0.273) & 0.212 (0.251) & 0.212 (0.244) \\ \hline
\multirow{8}{*}{\textbf{D}} & \multirow{4}{*}{200} & 0.1 & 0.016 (0.019) & 0.037 (0.052) & 0.043 (0.059) & 0.043 (0.059) \\
                            &                      & 0.3 & 0.055 (0.064) & 0.082 (0.105) & 0.102 (0.129) & 0.103 (0.124) \\
                            &                      & 0.5 & 0.108 (0.125) & 0.152 (0.190) & 0.174 (0.197) & 0.174 (0.191) \\
                            &                      & 0.7 & 0.187 (0.218) & 0.301 (0.474) & 0.283 (0.310) & 0.278 (0.298) \\\cline{2-7}
                            & \multirow{4}{*}{400} & 0.1 & 0.011 (0.014) & 0.029 (0.037) & 0.035 (0.044) & 0.034 (0.045) \\
                            &                      & 0.3 & 0.036 (0.046) & 0.066 (0.077) & 0.082 (0.102) & 0.083 (0.098) \\
                            &                      & 0.5 & 0.071 (0.089) & 0.118 (0.141) & 0.138 (0.165) & 0.135 (0.159) \\
                            &                      & 0.7 & 0.123 (0.155) & 0.233 (0.287) & 0.222 (0.259) & 0.218 (0.241) \\ 
\hline
\hline
\end{tabular}
\caption{Simulation results expressed in terms of $IMAE$ and the dispersion measure in brackets for the estimation of $m_\tau(\bx)$ with 30\% of censoring. The number of repetitions operated is $B=500$ for sample sizes $n \in \{200,400\}$ and with quantiles of interest $\tau \in \{0.1,0.3,0.5,0.7\}$.}
\label{table:model 1 censoring 30}
\end{table}

\begin{table}[t!]
\centering
\begin{tabular}{ccc||c|c|c|c}
\multicolumn{7}{c}{Censoring = 50\%} \\
DGP & $n$ & $\tau$ & $\widehat{m}_{cox,\, \tau}$ & $\widehat{m}_{si,\, \tau}$ & $\widehat{m}_{cop,\, \tau}^{(\independent)}$ & $\widehat{m}_{cop,\, \tau}^{(cox)}$\\
\hline
\hline
\multirow{8}{*}{\textbf{C}} & \multirow{4}{*}{200} & 0.1 & 0.019 (0.022)& 0.040 (0.056)& 0.044 (0.056)& 0.044 (0.054)\\
                            &                      & 0.3 & 0.063 (0.074)& 0.089 (0.111)& 0.113 (0.127)& 0.115 (0.127)\\
                            &                      & 0.5 & 0.123 (0.144)& 0.176 (0.242)& 0.201 (0.209)& 0.205 (0.213)\\
                            &                      & 0.7 & 0.213 (0.251)& 0.411 (0.402)& 0.354 (0.326)& 0.362 (0.326)\\\cline{2-7}
                            & \multirow{4}{*}{400} & 0.1 & 0.012 (0.015)& 0.031 (0.040)& 0.035 (0.042)& 0.035 (0.042)\\
                            &                      & 0.3 & 0.042 (0.051)& 0.071 (0.082)& 0.089 (0.106)& 0.089 (0.106)\\
                            &                      & 0.5 & 0.081 (0.099)& 0.136 (0.170)& 0.161 (0.179)& 0.163 (0.178)\\
                            &                      & 0.7 & 0.140 (0.171)& 0.299 (0.315)& 0.284 (0.290)& 0.281 (0.279)\\ \hline
\multirow{8}{*}{\textbf{D}} & \multirow{4}{*}{200} & 0.1 & 0.019 (0.022)& 0.040 (0.061)& 0.046 (0.059)& 0.044 (0.053)\\
                            &                      & 0.3 & 0.064 (0.074)& 0.090 (0.106)& 0.117 (0.129)& 0.115 (0.123)\\
                            &                      & 0.5 & 0.125 (0.143)& 0.202 (0.374)& 0.212 (0.218)& 0.210 (0.206)\\
                            &                      & 0.7 & 0.217 (0.249)& 0.526 (0.854)& 0.380 (0.341)& 0.380 (0.324)\\\cline{2-7}
                            & \multirow{4}{*}{400} & 0.1 & 0.013 (0.015)& 0.029 (0.038)& 0.038 (0.048)& 0.036 (0.045)\\
                            &                      & 0.3 & 0.043 (0.052)& 0.069 (0.082)& 0.099 (0.113)& 0.095 (0.105)\\
                            &                      & 0.5 & 0.084 (0.102)& 0.140 (0.173)& 0.177 (0.201)& 0.172 (0.174)\\
                            &                      & 0.7 & 0.145 (0.176)& 0.346 (0.632)& 0.306 (0.322)& 0.298 (0.283)\\
\hline
\hline
\end{tabular}
\caption{Simulation results expressed in terms of $IMAE$ and the dispersion measure in brackets for the estimation of $m_\tau(\bx)$ with 50\% of censoring. The number of repetitions operated is $B=500$ for sample sizes $n \in \{200,400\}$ and with quantiles of interest $\tau \in \{0.1,0.3,0.5,0.7\}$.}
\label{table:model 1 censoring 50}
\end{table}

\begin{figure}[t!]
	\begin{center}
	$\tau=0.5$
 \subfloat[$\widehat{m}_{si,\, \tau}$]{\includegraphics[scale=0.25]{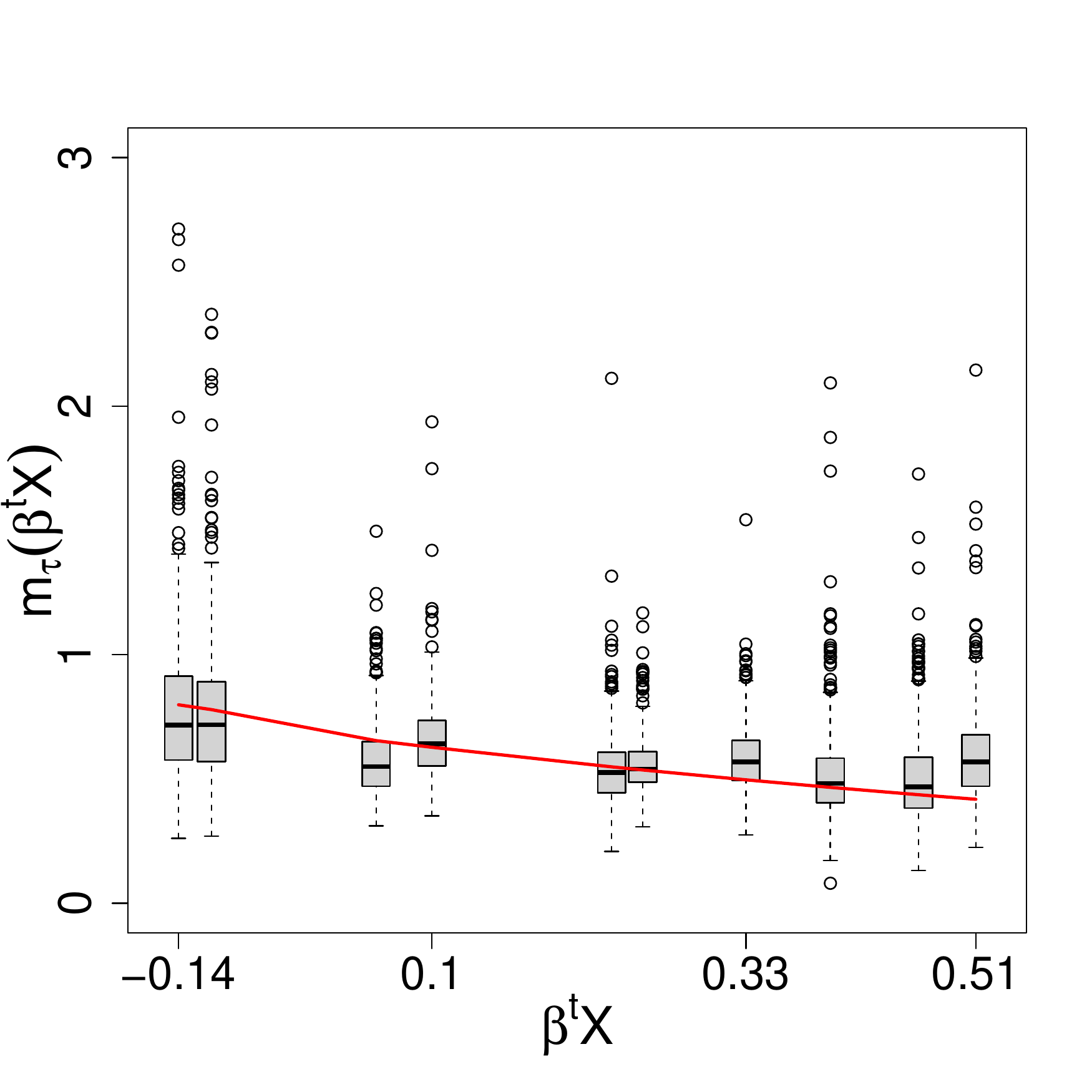}\label{im1SI}} 
 \qquad
 \subfloat[$\widehat{m}_{cop,\, \tau}^{(\independent)}$]{\includegraphics[scale=0.25]{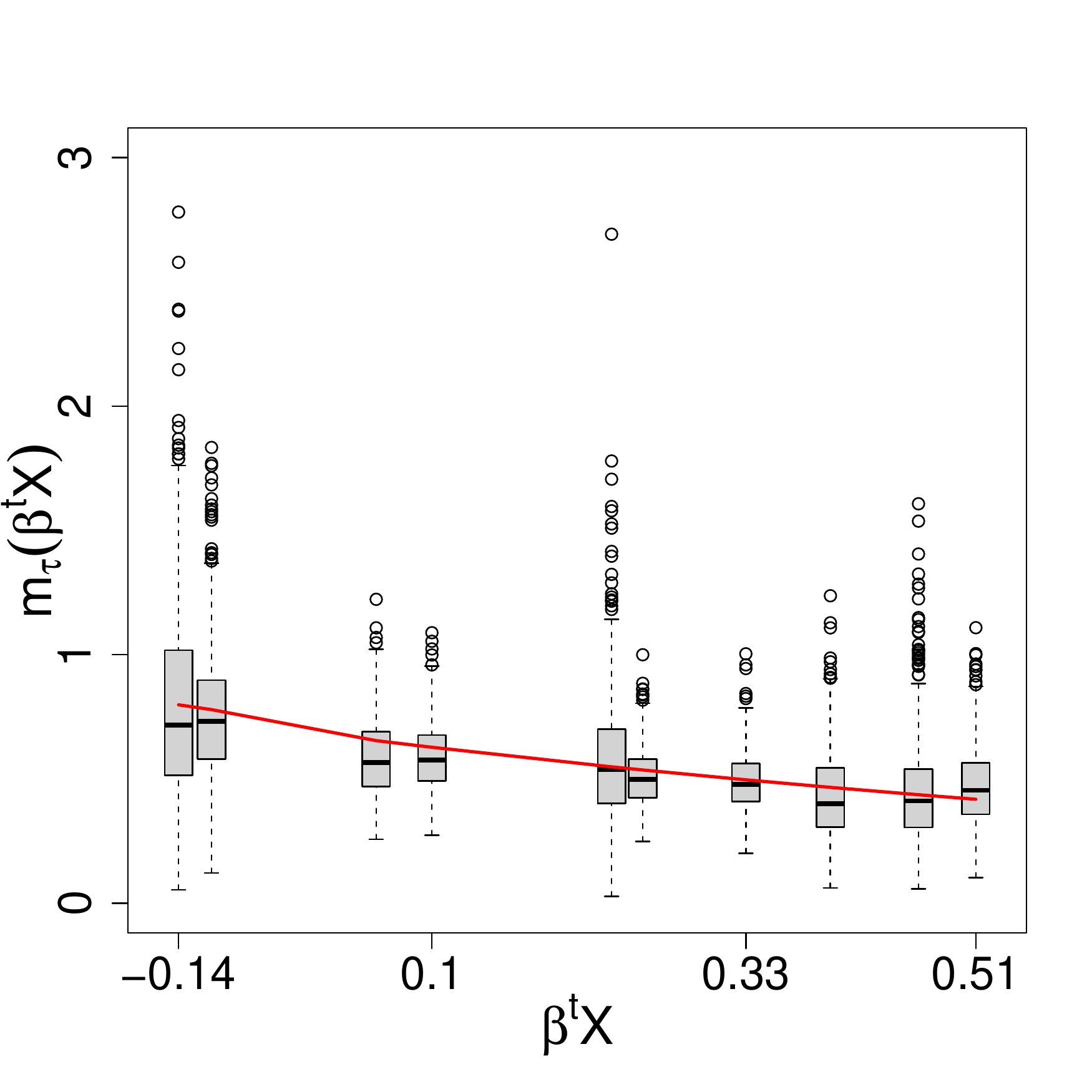}\label{im1COP_IND}} 
 \qquad
 \subfloat[$\widehat{m}_{cop,\, \tau}^{(cox)}$]{\includegraphics[scale=0.25]{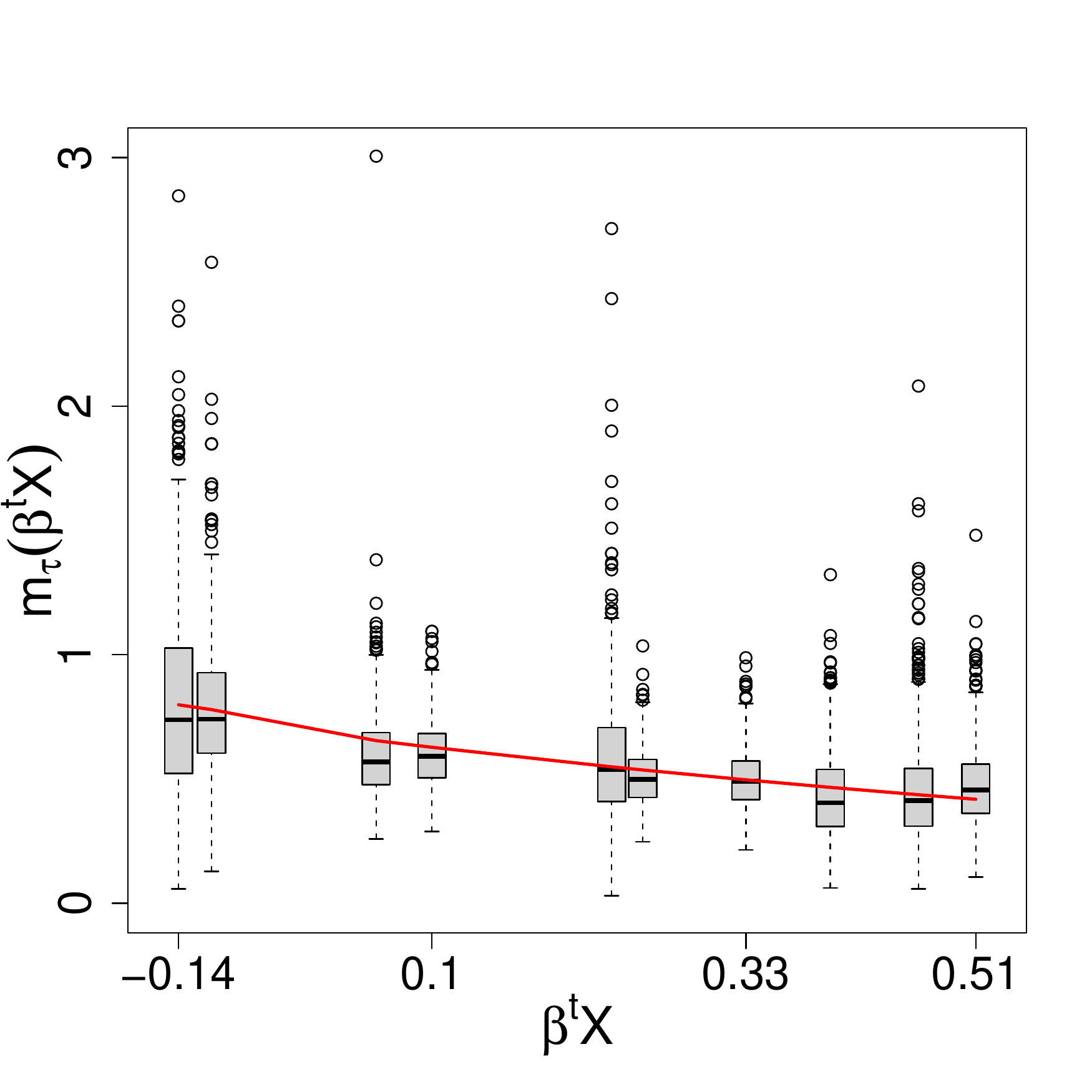}\label{im1COP_COX}}

	$\tau=0.7$
\subfloat[$\widehat{m}_{si,\, \tau}$]{\includegraphics[scale=0.25]{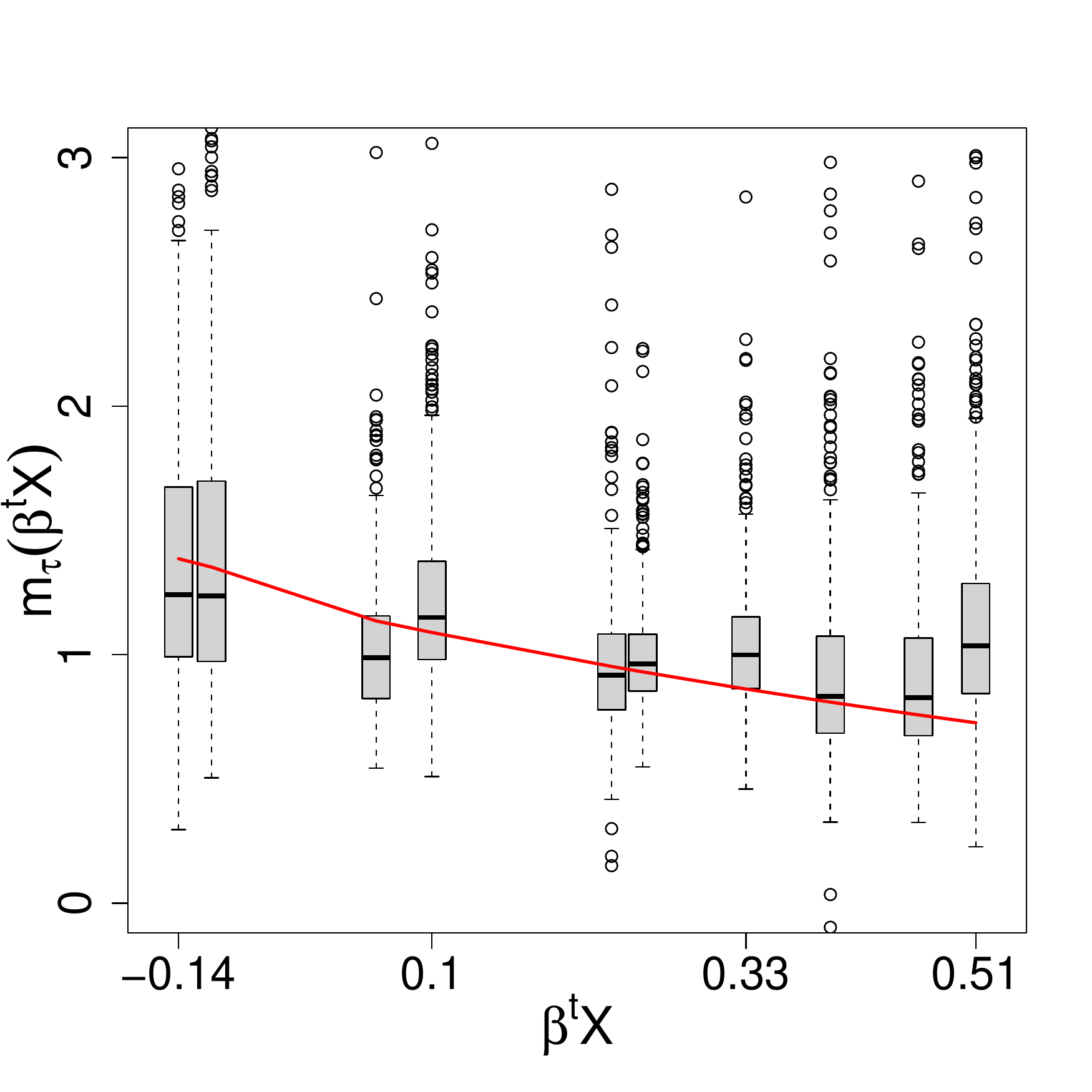}\label{im2SI}}
 \qquad
 \subfloat[$\widehat{m}_{cop,\, \tau}^{(\independent)}$]{\includegraphics[scale=0.25]{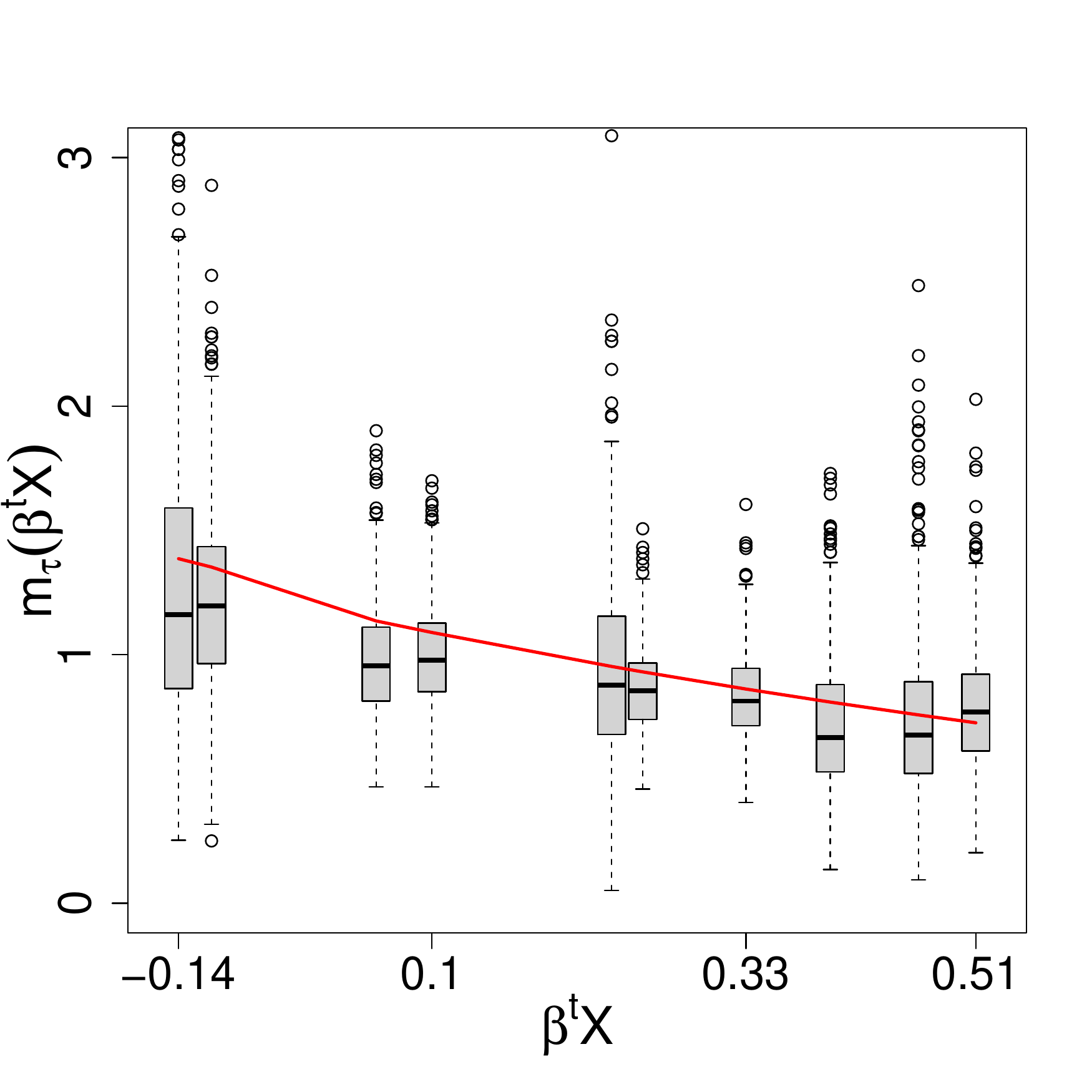}\label{im2COP_IND}}
 \qquad
 \subfloat[$\widehat{m}_{cop,\, \tau}^{(cox)}$]{\includegraphics[scale=0.25]{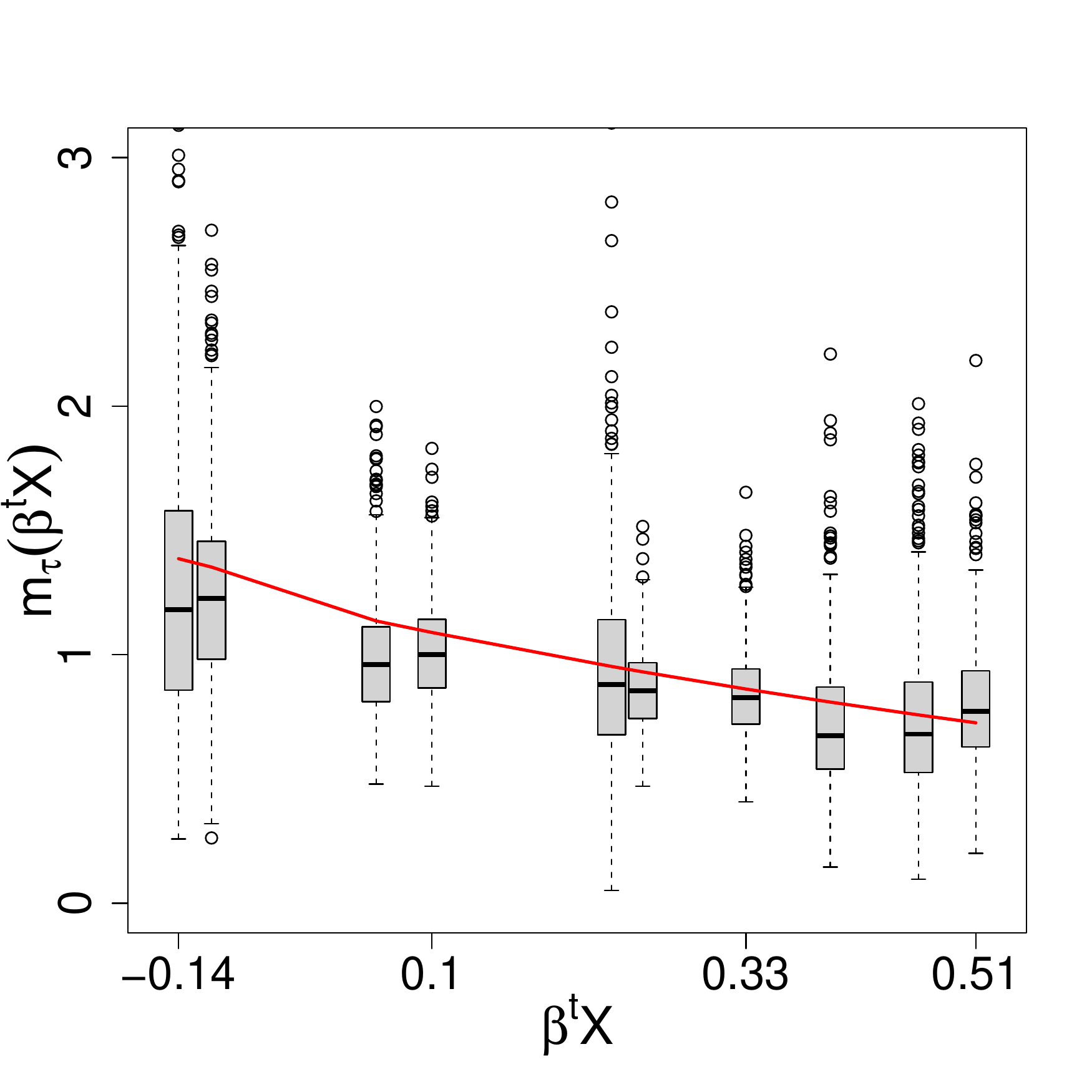}\label{im2COP_COX}}
\end{center}
\caption{Boxplots of three estimators considered for $m_\tau(\bx)$ for $N=10$ points spread on the domain of $\beta^{\mathsf{T}}\bX^{(1)}$ in DGP \textbf{C}, and for $\tau \in \{0.5,0.7\}$. All pictures are based on $B=500$ repetitions for $n=400$ and $30\%$ censoring. The red lines represent the true value of $m_{0.5}(\bx)$ and $m_{0.7}(\bx)$.}
	\label{fig:setup_1_n400_tau0507}
\end{figure}

Figure \ref{fig:setup_1_n400_tau0507} serves to illustrate these results for the particular case of DGP \textbf{C} with $n=400$ and for high quantile levels $(\tau \in \{0.5,0.7\})$. For the sake of brevity, we only report here a graphical comparison between the copula-based estimators and the single-index estimator, as $\widehat{m}_{cox,\, \tau}$ clearly outperforms the latter. As can be observed from Figure \ref{fig:setup_1_n400_tau0507}, the difference between $\widehat{m}_{si,\, \tau}$ and both copula-based estimators is graphically modest, especially for the highest quantile of interest $\tau=0.7$, hereby reinforcing the previously discussed results of Tables \ref{table:model 1 censoring 30} and \ref{table:model 1 censoring 50}. Furthermore, note that all estimators tend to present worse performances for low levels of $\beta^{\mathsf{T}}\bx$. This is expected in this simulation setting, as these levels correspond to the region of $\beta^{\mathsf{T}}\bx$ where the exact conditional censoring probability is the highest.

Focusing again on Tables \ref{table:model 1 censoring 30} and \ref{table:model 1 censoring 50}, we further note that, while there is logically a difference between the performances of $\widehat{m}_{cop,\, \tau}^{(\independent)}$ and $\widehat{m}_{cop,\, \tau}^{(cox)}$ depending on the simulated scenario, the effect of an appropriate modelling of $\widehat{G}_C(\cdot|\bx)$ seems to be rather limited in this simulation setup for our estimators. Additionally, while both $\widehat{m}_{si,\, \tau}$ and $\widehat{m}_{cop,\, \tau}^{(\independent)}$ rely here on the assumption of independence between the censoring variable and the covariate vector, the latter seems to numerically behave better when confronted to the violation of this hypothesis. This can be observed from the comparison between DGP \textbf{C} and \textbf{D} of the dispersion measures of both estimators, once again especially for high levels of quantile values and censoring percentages. Of course, one could argue that, given the results for the dispersion measure of $\widehat{m}_{si,\, \tau}$ for high quantiles, this can partly be due to a poor smoothing parameter choice. However, the latter was implemented using the proposed methodology of \citet{BEGVK14}, just as a practitioner would have resolved to act. 


\newpage 
\subsubsection[Model 2]{Model 2: covariate vector $\bX^{(2)}$}
To complete our simulation study, we now consider in this section the second model where the time-to-event data is simulated using covariate vector $\bX^{(2)}$. Consequently, the resulting quantile regression is no longer a single-index nor a Cox regression model in $(X_1, X_2, \ldots, X_5)$. For the sake of brevity, we only consider here the situation where the censoring variable is independent from the covariate vector, as the impact of a dependent scheme has been treated in model 1. Specifically, we simulate the censoring variable from an exponential distribution with parameter values $0.208$ and $0.486$ for approximately 30\% and 50\% censoring. For comparison purposes, we consider the four estimation procedures described in model 1, given covariate vector $(X_1, X_2, \ldots, X_5)$. The copula-based estimation procedures are constructed using the same semiparametric modelling strategy as for model 1. 

The resulting performance of the considered estimators are depicted in Table \ref{table:model 2}, once again in terms of $IMAE$. In this simulation scheme, we observe that both copula-based estimators tend to outperform their competitors, with the exception of estimations for very low quantiles of interest, where the effect of a (`small') misspecification of the underlying model seems to be moderate for both $\widehat{m}_{cox,\, \tau}$ and $\widehat{m}_{si,\, \tau}$. As we move away from these low quantile levels, the consequences of misspecifying the model become more severe for the competing estimators, notably for a purely parametric approach ($\widehat{m}_{cox,\, \tau}$). In contrast, the copula-based approach presents satisfactory results for varying censoring proportions and sample sizes, especially when keeping in mind that the simulated scenario is `only' a slightly altered version of the ideal scenario for its competitors. Conclusively, this advocates, here again, for the flexibility of our procedure and its withstanding to misspecification of the underlying model for multidimensional problems when comparing with other semiparametric or fully parametric modelling techniques.

\begin{table}[ht!]
\centering
\begin{tabular}{ccc||c|c|c|c}
\multicolumn{7}{c}{Model 2} \\
$n$ & $p_c$ & $\tau$ & $\widehat{m}_{cox,\, \tau}$ & $\widehat{m}_{si,\, \tau}$ & $\widehat{m}_{cop,\, \tau}^{(\independent)}$ & $\widehat{m}_{cop,\, \tau}^{(cox)}$\\
\hline
\hline
\multirow{8}{*}{200} & \multirow{4}{*}{0.3} & 0.1 & 0.115 (0.046) & 0.105 (0.142) & 0.115 (0.160) & 0.115 (0.159) \\
                            &               & 0.3 & 0.391 (0.156) & 0.279 (0.273) & 0.249 (0.285) & 0.250 (0.286) \\
                            &               & 0.5 & 0.760 (0.302) & 0.489 (0.538) & 0.452 (0.532) & 0.454 (0.538) \\
                            &               & 0.7 & 1.319 (0.525) & 0.988 (1.023) & 0.781 (0.828) & 0.783 (0.842) \\\cline{2-7}
                     & \multirow{4}{*}{0.5} & 0.1 & 0.116 (0.054) & 0.107 (0.156) & 0.115 (0.139) & 0.116 (0.143) \\
                            &               & 0.3 & 0.391 (0.182) & 0.334 (0.324) & 0.303 (0.335) & 0.311 (0.338) \\
                            &               & 0.5 & 0.760 (0.359) & 0.611 (0.742) & 0.574 (0.552) & 0.583 (0.558) \\
                            &               & 0.7 & 1.321 (0.615) & 1.441 (1.142) & 1.053 (0.845) & 1.074 (0.865) \\ \hline
\multirow{8}{*}{400} & \multirow{4}{*}{0.3} & 0.1 & 0.115 (0.029) & 0.084 (0.101) & 0.092 (0.118) & 0.092 (0.120) \\
                            &               & 0.3 & 0.388 (0.099) & 0.219 (0.222) & 0.211 (0.264) & 0.212 (0.265) \\
                            &               & 0.5 & 0.753 (0.192) & 0.372 (0.401) & 0.358 (0.427) & 0.359 (0.430) \\
                            &               & 0.7 & 1.308 (0.333) & 0.685 (0.844) & 0.598 (0.682) & 0.604 (0.683) \\\cline{2-7}
                     & \multirow{4}{*}{0.5} & 0.1 & 0.114 (0.035) & 0.084 (0.105) & 0.096 (0.109) & 0.096 (0.112) \\
                            &               & 0.3 & 0.387 (0.118) & 0.254 (0.231) & 0.221 (0.283) & 0.226 (0.285) \\
                            &               & 0.5 & 0.751 (0.228) & 0.479 (0.490) & 0.431 (0.486) & 0.438 (0.488) \\
                            &               & 0.7 & 1.305 (0.397) & 0.933 (0.878) & 0.850 (0.781) & 0.857 (0.793) \\ 
\hline
\hline
\end{tabular}
\caption{Simulation results expressed in terms of $IMAE$ and the dispersion measure in brackets for the estimation of $m_\tau(\bx)$ for model 2. The number of repetitions operated is $B=500$ for sample sizes $n \in \{200,400\}$, censoring proportion $p_c \in \{0.3,0.5\}$ and with four levels of quantile of interest $\tau \in \{0.1,0.3,0.5,0.7\}$.}
\label{table:model 2}
\end{table}

\section{Real Data Application}\label{section:application}
We present in this section a brief application of our procedure by analysing the Colorado Plateau uranium miners cohort data (see e.g. \citet{L95}, \citet{LG96}). The object of the study, for which 3347 Caucasian male miners having worked at least a month in the uranium mines of the Colorado Plateau were followed, is to investigate the risk of lung cancer related to smoking and radon exposure. Hence, the event of interest is defined as the time till lung cancer death (expressed as the logarithm of number of years), which affected a total of 258 miners. Besides failure time, the study also includes information about age at entry to the study, cumulative smoking (in number of packs) and radon exposure (in working level month (WLM)). 

\begin{figure}[t!]
	\begin{center}
 \subfloat{\includegraphics[scale=0.274]{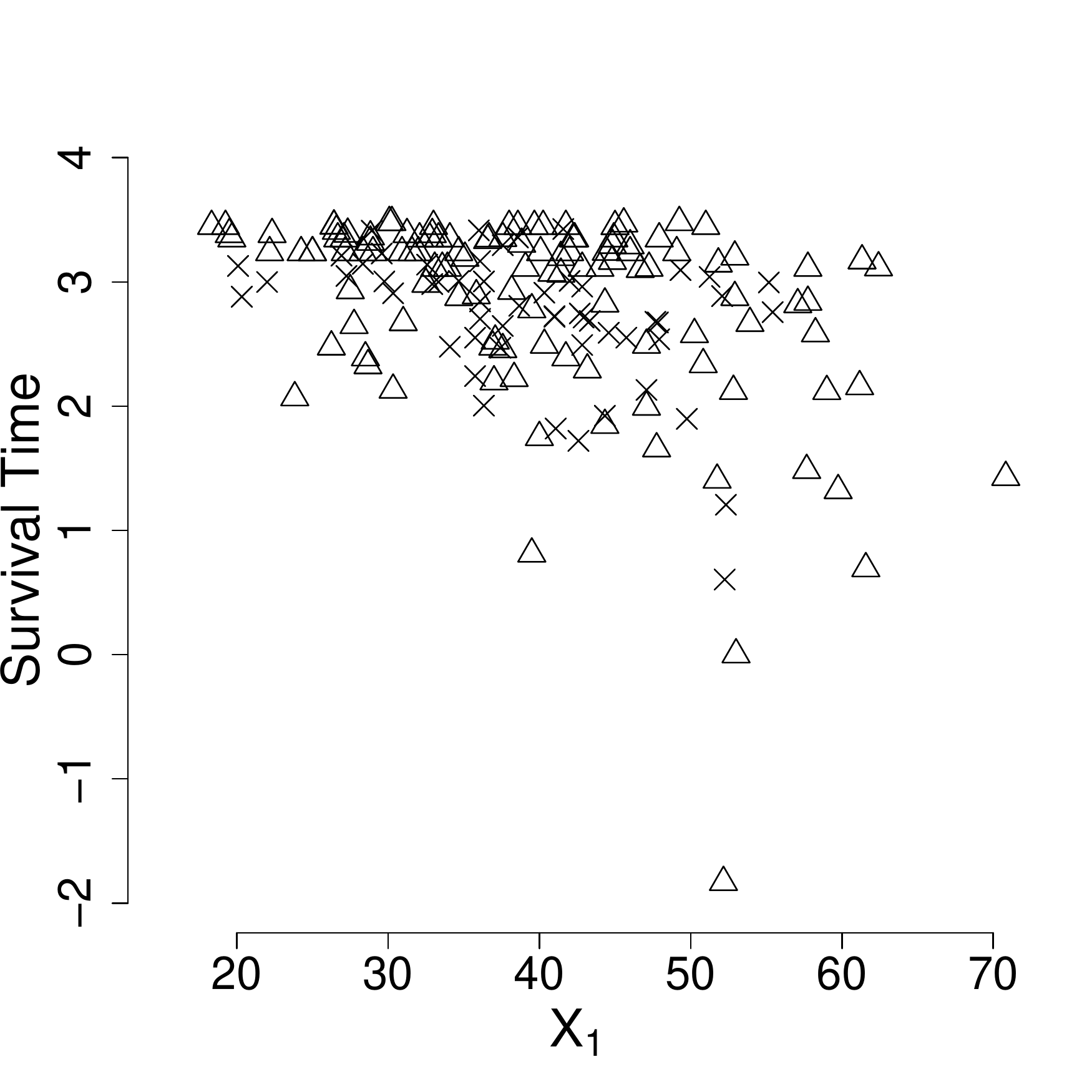}}
 \qquad
 \subfloat{\includegraphics[scale=0.274]{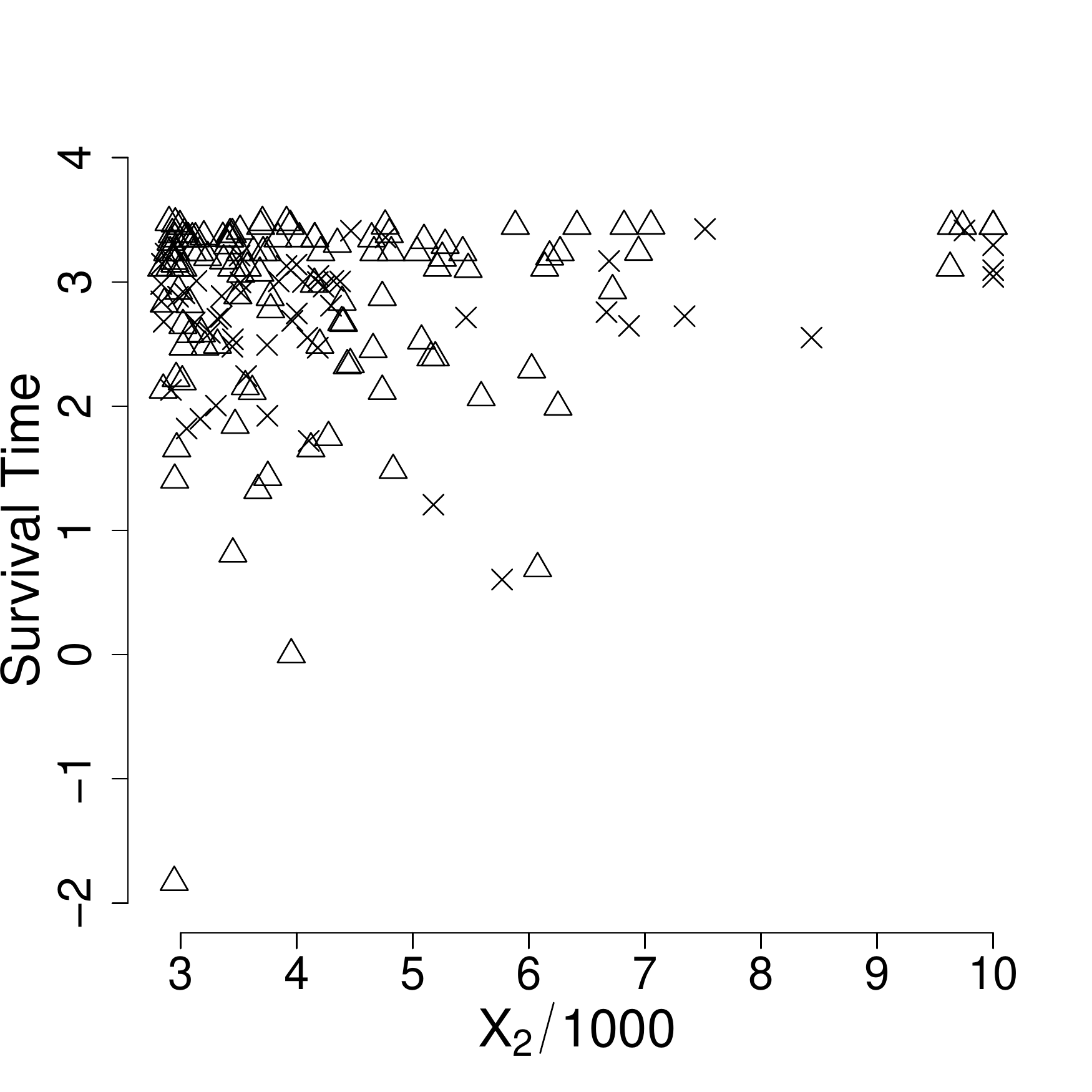}}
 \qquad
 \subfloat{\includegraphics[scale=0.274]{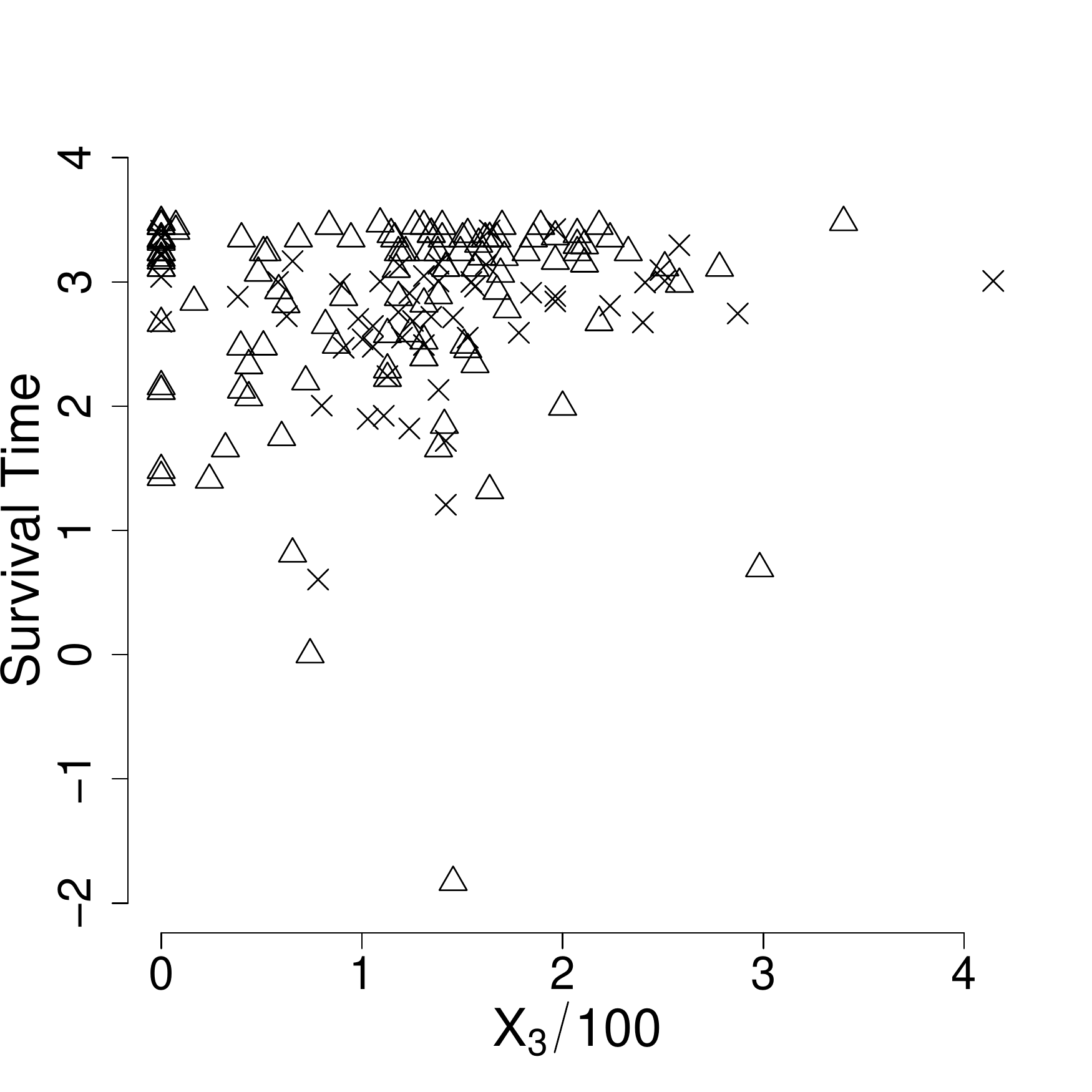}}
	\end{center}
	\caption{Colorado Plateau uranium miners cohort data. Scatter plots of the survival time versus each covariate. Uncensored data points are given by $\times$, while censored observations are represented by $\bigtriangleup$.}
	\label{fig: application scatter}
\end{figure}

As the original data set is prone to heavy censoring (92.3\%), and given the illustrative nature of this section, we first define a subsample on which the analysis will be performed. To that end, in order to preserve as best as possible the nature of the population at risk, we decide to define a threshold on the radon exposure above which observations will enter the subsample. The value of the threshold is practically chosen as a trade-off between censoring proportion and actual number of observations that are to be selected. Specifically, we find that by defining a threshold of 2831 WLM on radon exposure, a subsample of 176 observations is constituted, 55 of which were subject to the event of interest. Scatterplots of the selected data are represented in Figure \ref{fig: application scatter}, where $X_1$ is the age at entry into the study, $X_2$ is the cumulative radon exposure and $X_3$ is the cumulative smoking. 

Regarding the data analysis, endorsing the role of a practitioner, we are faced with the choice of an appropriate estimator for the application of quantile regression in this context. We therefore consider the following distinctive candidates:

\begin{enumerate}
	\item[$\widehat{m}_{cop,\, \tau}^{(cox)}$]: Copula-based quantile regression estimator, with Cox regression modelling  for $G_C(\cdot|\cdot)$.  
	\item[$\widehat{m}_{si, \, \tau}$]: Single-Index methodology of \citet{BEGVK14}. 
	\item[$\widehat{m}_{cox, \,\tau}$]: Semiparametric estimator based on the Cox proportional hazards model. Specifically, $\widehat{m}_{cox, \tau}(\bx) = \widehat{H}_0^{-1}(-\log(1-\tau)\exp(-\widehat{\beta}^{\mathsf{T}} \bx))$, where $\widehat{\beta}$ is estimated by maximum partial likelihood, and $\widehat{H}_0$ is the Nelson-Aalen-type estimator of the cumulative baseline hazard. 
\end{enumerate}

As a general evaluation measure to compare models for the present data set, we consider the median quantile loss from predicting the $\tau$-th conditional quantile of $T$ for the uncensored observations. In other words, we use the following cross-validated prediction error criterion:
$$ \text{PE}\left(\widehat{m}_\tau\right) = \text{med}_{\stackrel{1\leq i\leq n}{\Delta_i=1}}\; \rho_\tau\left(Y_i - \widehat{m}_\tau^{-i}(\bX_i) \right),$$
where $\widehat{m}_\tau^{-i}$ denotes any estimator of $m_\tau$ based on all observations except the $i$-th one. 

For the implementation of the copula-based regression estimator, we adopt the methodology of Section \ref{sect:background} and our simulation study by opting for a semiparametric modelling of the four-variate copula density, where the bivariate copulas of the response variable with each covariate are estimated using the procedure of \citet{GCP14} with quadratic polynomials along with the proposed data-driven bandwidth selection scheme. The remaining trivariate noisy copula is, afterwards, estimated using vine techniques with the same candidate families and selection criterion as in Section \ref{section:simulations}. Additionally, a general appropriate hypothesis has to be made for the modelling of the conditional distribution $G_C(\cdot|\cdot)$. As it is shown in the literature that independence is unsuitable for the data of the study (see e.g. \citet{LT13}), we propose to model the conditional distribution through a semiparametric Cox regression for the censoring time with respect to the covariates. Therefore, as is common in practice, the parameter of the regression is estimated by maximum partial likelihood while the Breslow estimator is used for computing the baseline survivor function. 

Concerning $\widehat{m}_{si, \tau}$, the bandwidth choice is performed using the proposed leave-one-out cross-validation procedure on normalized covariates with 15 candidates ($h \in \{0.1,0.2,\ldots,1.5\}$). Note that, as the latter procedure is computationaly costly, it is here assumed that the selected bandwidth is constant for all datasets entering the estimation of $\text{PE}(\widehat{m}_{si, \tau})$, which is of little impact in this context given that only one observation at the time is to be removed from each dataset. 

The results of the evaluation measure are reported in Table \ref{table:application CV error} for four quantile levels of interest $\tau \in \{0.1,0.3,0.5,0.7\}$. It is observed that, in terms of cross-validated prediction error, our copula-based approach depicts quite confidently the best performance for every considered quantile level of interest. By comparison, the single-index structure seems to be inappropriate as such for the studied data set, and should possibly require further work and attention on, for instance, transformations of covariates. Furthermore, its sensitivity to higher quantile levels is, here again, highlighted. Lastly, despite being part of every practitioner's toolbox for survival analysis, the proportional hazards regression estimator is also relatively largely outperformed by the increased flexibility of the copula-based approach. Hence, this example clearly illustrates the ability of our estimator to adapt to the underlying regression structure of the data. 

\begin{table}[ht!]
\centering
\begin{tabular}{c||c|c|c}
$\tau$ & $\text{PE}\left(\widehat{m}_{cop,\, \tau}^{(cox)}\right)$ & $\text{PE}\left(\widehat{m}_{si,\, \tau}\right)$ & $\text{PE}\left(\widehat{m}_{cox,\, \tau}\right)$ \\
\hline
\hline
0.1 &  0.348 & 0.848 & 0.946\\
0.3 &  0.774 & 1.471 & 1.180\\
0.5 &  0.785 & 1.901 & 1.169\\
0.7 &  0.470 & 2.838 & 0.967\\
\hline
\hline
\end{tabular}
\caption{Colorado Plateau uranium miners cohort data. Prediction error multiplied by 10 for each censored quantile regression estimator for quantile levels $\tau \in \{0.1,0.3,0.5,0.7\}$.}
\label{table:application CV error}
\end{table}

Conclusively, recalling that, from equation \eqref{eq:Cop CQR prop mono}, our procedure enjoys the additional valuable property of being automatically monotonic across quantile levels, this real data application plainly highlights the relevance of our estimator for flexible analyses of multivariate censored data.

\section{Conclusion}

In this work we have proposed a semiparametric copula-based quantile regression estimator in the context of potentially right-censored responses. On a general note, for data with or without censoring, and motivated by the regression context, a novel semiparametric estimation approach for the implied copula density was studied. Furthermore, in parallel to the procedure of \citet{NEGVK15}, the proposed regression estimator in this work is obtained as a weighted quantile of the observed response variable, hereby opening the door to the practical use of the quantile regression code developed by \citet{KP97} and \citet{K05}. Asymptotic normality of the resulting estimator for both complete and censored data was obtained with convergence rate determined by the nonparametric estimation of bivariate copula densities. Finally, supporting the objective of this work, an extensive simulation study and a real data application have been carried out to illustrate both the validity of the semiparametric copula estimation and the increased flexibilty of our procedure in comparison with existing alternatives, especially when no a priori knowledge about the functional form of the quantile regression is available in practice.  

\section*{Acknowledgments}
All authors acknowledge financial support from IAP research network P7/06 of the Belgian Government (Belgian Science Policy) and from the contract `Projet d’Actions de Recherche Concert\'{e}es' (ARC) 11/16-039 of the `Communaut\'{e} fran\c caise de Belgique', granted by the `Acad\'{e}mie universitaire Louvain'.

Computational resources have been provided by the supercomputing facilities of the Universit\'{e} catholique de Louvain (CISM/UCL) and the Consortium des \'{E}quipements de Calcul Intensif en F\'{e}d\'{e}ration Wallonie Bruxelles (C\'{E}CI) funded by the Fonds de la Recherche Scientifique de Belgique (F.R.S.-FNRS) under convention 2.5020.11.

\newpage
\section*{Appendix}\label{section:appendix}
\renewcommand{\theequation}{A.\arabic{equation}}
We develop in this appendix the proofs of Theorem \ref{theorem1} and Corollary \ref{corr1}. 
\subsection*{Proof of Theorem \ref{theorem1}}
\begin{proof}[\nopunct]
Define 
$$
\widehat{A}_n(s)=\sum_{i=1}^n \left[\rho_{\tau}(\epsilon_i-s/a_n)-\rho_{\tau}(\epsilon_i)\right]\widehat{W}_i(\bx) \, \widehat{c}^{u}_{Y\bX}\big(\widehat{F}^{u}_Y(Y_i), \widehat{\bF}^{u}(\bx)\big),
$$
with $\epsilon_i\equiv \epsilon_i(\bx,\tau)=Y_i-m_{\tau}(\bx),$ and $a_n=\sqrt{nh^2}$. Observe first that, by definition of $\widehat{m}_{\tau}(\bx)$, 
$$
a_n(\hat{m}_{\tau}(\bx)-m_{\tau}(\bx))=\arg\min_s\widehat{A}_n(s).
$$
Furthermore, given that $\rho_{\tau}$ is a convex function and that $\widehat{W}_i(\bx) \, \widehat{c}^{u}_{Y\bX}\big(\widehat{F}^{u}_Y(Y_i), \widehat{\bF}^{u}(\bx)\big) \geq 0$ for all $i=1,\ldots,n$, we have that $s \mapsto \widehat{A}_n(s)$ is convex. The idea is then to develop an expression of $\widehat{A}_n(s)$ leading to the application of the quadratic approximation Lemma of convex functions (Basic Corollary in \citet{HP93}). To that end, we have that (Knight's (1998) identity) 
$$
\rho_{\tau}(u-v)-\rho_{\tau}(u)=-v\psi_{\tau}(u)+R(u,v),
$$ 
with  $\psi_{\tau}(u)=\tau-I(u \le 0)$,
\begin{align*}
R(u,v)&=\int_0^v(I(u\le s)-I(u\le 0))ds=(u-v)(I(u\le 0)-I(u\le v)), 
\end{align*}
and $0 \le R(u,v) \le |v|$. Hence, we may write 
$$
\widehat{A}_n(s)=-s\widehat{A}_{1n}+\hat{A}_{2n}(s),
$$
with 
$$
\widehat{A}_{1n}= a_n^{-1} \sum_{i=1}^n \psi_{\tau}(\epsilon_i)\,\widehat{W}_i(\bx)\,\widehat{c}^{u}_{Y\bX}\big(\widehat{F}^{u}_Y(Y_i), \widehat{\bF}^{u}(\bx)\big),
$$
and 
$$
\widehat{A}_{2n}(s)=\sum_{i=1}^n R(\epsilon_i,s/a_n)\,\widehat{W}_i(\bx)\,\widehat{c}^{u}_{Y\bX}\big(\widehat{F}^{u}_Y(Y_i), \widehat{\bF}^{u}(\bx)\big).
$$
Focusing first on $\widehat{A}_{2n}(s)$, we will show that 
\begin{align}\label{eq:A2n_hat} 
\widehat{A}_{2n}(s)= \frac{s^2}{2}w^{-1}(\bx)f_{T|\bX}(m_{\tau}(\bx)|\bx) \, \frac{n}{a_n^2}+o_p\left(\frac{n}{a_n^2}\right),
\end{align}
where $f_{T|\bX}$ is the conditional density of $T$ given $\bX$ and $w(\bx) = p(\bx)/\big[\PP(\Delta=1)c^u_{\bX}(\bF^u(\bx))\big]$. To that end, we write
$$
\widehat{A}_{2n}(s)= A_{2n}(s) + \Delta_{1n}(s) + \Delta_{2n}(s),
$$
where
\begin{align*}
A_{2n}(s)&=\sum_{i=1}^n R(\epsilon_i,s/a_n)\,W_i(\bx)\,c^{u}_{Y\bX}\big(F^{u}_Y(Y_i), \bF^{u}(\bx)\big)\\
\Delta_{1n}(s)&=\sum_{i=1}^n R(\epsilon_i,s/a_n)\,(\widehat{W}_i(\bx)-W_i(\bx))\,c^{u}_{Y\bX}\big(F^{u}_Y(Y_i), \bF^{u}(\bx)\big)\\
\Delta_{2n}(s)&=\sum_{i=1}^n R(\epsilon_i,s/a_n)\,\widehat{W}_i(\bx)\,\big[\widehat{c}^{u}_{Y\bX}\big(\widehat{F}^{u}_Y(Y_i), \widehat{\bF}^{u}(\bx)\big)-c^{u}_{Y\bX}\big(F^{u}_Y(Y_i), \bF^{u}(\bx)\big)\big].
\end{align*} 
Concentrating on each term, we will show that
\begin{align}
A_{2n}(s)&=\frac{s^2}{2}w^{-1}(\bx)f_{T|\bX}(m_{\tau}(\bx)|\bx) \, \frac{n}{a_n^2}+o_p\left(\frac{n}{a_n^2}\right) \label{eq:A2n} \\
\Delta_{1n}(s)&=o_p\left(\frac{n}{a_n^2}\right) \label{eq:Delta_1n}\\
\Delta_{2n}(s)&=o_p\left(\frac{n}{a_n^2}\right) \label{eq:Delta_2n}.
\end{align}
For the proof of \eqref{eq:A2n}, we first establish that
\begin{align*}
\E(A_{2n}(s))&=n\,\E\left(R(\epsilon_1,s/a_n)\,W_1(\bx)\,c^{u}_{Y\bX}\big(F^{u}_Y(Y_1), \bF^{u}(\bx)\big)\right)\\
& =  n\,w^{-1}(\bx) \int_{0}^{s/a_n}\big(F_{T|\bX}(m_{\tau}(\bx)+t|\bx)-F_{T|\bX}(m_{\tau}(\bx)|\bx)\big)\mathrm dt, 
\end{align*}
where $F_{T|\bX}$ denotes, as in Section \ref{sect:background}, the conditional c.d.f. of $T$ given $\bX$, and where, for the second equality, we used (see Section \ref{subsect:our proc}) the fact that, for any measurable function $\varphi: \RR \rightarrow \RR$, we have $\E(\Delta\varphi(Y)\, c^{u}_{Y\bX}\big(F^{u}_Y(Y), \bF^{u}(\bx)\big))=w^{-1}(\bx)\E(\Delta\varphi(Y)|\bX=\bx)$. Next, we write 
\begin{align*}
F_{T|\bX}(m_{\tau}(\bx)+t|\bx)-F_{T|\bX}(m_{\tau}(\bx)|\bx)&=t \, f_{T|\bX}(m_{\tau}(\bx)+\theta t|\bx)\\
&=t\,f_{T|\bX}(m_{\tau}(\bx)|\bx)+t\,R(t), 
\end{align*}
for some $\theta \in (0,1)$, with $R(t)=f_{T|\bX}(m_{\tau}(\bx)+\theta t|\bx)-f_{T|\bX}(m_{\tau}(\bx)|\bx).$ This yields
$$
\E(A_{2n}(s))=\frac{s^2}{2}w^{-1}(\bx)f_{T|\bX}(m_{\tau}(\bx)|\bx) \,\frac{n}{a_n^2}+n\,w^{-1}(\bx)\int_{0}^{s/a_n}t\,R(t)\mathrm dt.
$$
From the fact that, under the required bandwidth condition and assumption \ref{cond:true f},
$$
\left|\int_{0}^{s/a_n}t\,R(t)\mathrm dt\right| \le \frac{s^2}{2\, a_n^2}\sup_{|t|\le |s|/a_n}|R(t)|=o(1/a_n^2),
$$
we get that
$$
\E(A_{2n}(s))=\frac{s^2}{2}w^{-1}(\bx)f_{T|\bX}(m_{\tau}(\bx)|\bx) \, \frac{n}{a_n^2} +o\left(\frac{n}{a_n^2}\right),
$$ 
provided assumption \ref{cond:pt x} is satisfied. To conclude the proof of \eqref{eq:A2n}, it is then sufficient to show that 
$$
\V(A_{2n}(s))=o\left(\frac{n}{a_n^2}\right)^2.
$$  
To that end, observe that, for $n$ sufficiently large, and under assumptions \ref{cond:true f},\ref{cond:pt x} and \ref{cond:G_C},
\begin{align*}
\V(A_{2n}(s))&\le n \, \E\left(R(\epsilon_1,s/a_n)\,W_1(\bx)\,c^{u}_{Y\bX}\big(F^{u}_Y(Y_1), \bF^{u}(\bx)\big)\right)^2\\
& \le n \, \E\left(R(\epsilon_1,s/a_n)\,W_1(\bx)\,c^{u}_{Y\bX}\big(F^{u}_Y(Y_1), \bF^{u}(\bx)\big)\right) \frac{|s|}{a_n}\big(1-G_C(m_{\tau}(\bx)+\delta)|\bx\big)^{-1}\\
& \times \sup_{t\in \RR}c^{u}_{Y\bX}\big(F^{u}_Y(t), \bF^{u}(\bx)\big), \quad\text{for some $\delta >0$}\\
&= O\left(\frac{n}{a_n^3}\right)=o\left(\frac{n}{a_n^2}\right)^2,
\end{align*}
as $a_n/n$ converges to 0.

Concentrating now on the proof of \eqref{eq:Delta_1n}, observe that, for $n$ sufficiently large 
\begin{align*}
|\Delta_{1n}(s)|&=\left|\sum_{i=1}^{n}R(\epsilon_i,s/a_n)\,W_i(\bx)\,c^{u}_{Y\bX}\big(F^{u}_Y(Y_i), \bF^{u}(\bx)\big)\, \frac{\widehat{G}_C(Y_i-|\bx)-G_C(Y_i-|\bx)}{1-\widehat{G}_C(Y_i-|\bx)}\right|\\
& \le A_{2n}(s) \sup_{t\leq \max_{\stackrel{1\leq i\leq n}{\Delta_i=1}} Y_i}\frac{|\widehat{G}_C(t|\bx)-G_C(t|\bx)|}{1-\widehat{G}_C(t|\bx)}\\
&=o_p\left(\frac{n}{a_n^2}\right),
\end{align*}
provided that $\widehat{G}_C(\cdot|\bx)$ satisfies assumption \ref{cond:G_C_hat}.

Lastly, for the proof of \eqref{eq:Delta_2n}, note that, for $n$ sufficiently large, 
\begin{align*}
|\Delta_{2n}(s)|&=\left|\sum_{i=1}^{n}R(\epsilon_i,s/a_n)\,\widehat{W}_i(\bx)\,c^{u}_{Y\bX}\big(F^{u}_Y(Y_i), \bF^{u}(\bx)\big)\, \frac{\widehat{c}^{u}_{Y\bX}\big(\widehat{F}^{u}_Y(Y_i), \widehat{\bF}^{u}(\bx)\big)-c^{u}_{Y\bX}\big(F^{u}_Y(Y_i), \bF^{u}(\bx)\big)}{c^{u}_{Y\bX}\big(F^{u}_Y(Y_i), \bF^{u}(\bx)\big)}\right|\\
& \le (A_{2n}(s) + \Delta_{1n}(s)) \sup_{t\in \RR}\frac{\big|\widehat{c}^{u}_{Y\bX}\big(\widehat{F}^{u}_Y(t), \widehat{\bF}^{u}(\bx)\big)-c^{u}_{Y\bX}\big(F^{u}_Y(t), \bF^{u}(\bx)\big)\big|}{c^{u}_{Y\bX}\big(F^{u}_Y(t), \bF^{u}(\bx)\big)}\\
&=o_p\left(\frac{n}{a_n^2}\right),
\end{align*}
provided that assumption \ref{cond:cop_hat}-(i) is satisfied. 

Hence, reassembling \eqref{eq:A2n}, \eqref{eq:Delta_1n} and \eqref{eq:Delta_2n} yields 
$$
\frac{a_n^2}{n}\widehat{A}_n(s)=-s\frac{a_n^2}{n}\widehat{A}_{1n}+\frac{s^2}{2}w^{-1}(\bx)f_{T|\bX}(m_{\tau}(\bx)|\bx)+o_p(1).
$$ 
Therefore, by the quadratic approximation Lemma of a convex function, see e.g. \citet{HP93}, if $\frac{a_n^2}{n}\hat{A}_{1n}=O_p(1)$, then 
\begin{align}\label{eq:quadr_lemma}
a_n(\widehat{m}_{\tau}(\bx)-m_{\tau}(\bx))=\arg\min_s\frac{a_n^2}{n}\widehat{A}_n(s)=\frac{w(\bx)}{f_{T|\bX}(m_{\tau}(\bx)|\bx)}\frac{a_n^2}{n}\widehat{A}_{1n}+o_p(1).
\end{align}
As a consequence, the asymptotic behaviour of our estimator will be driven by the asymptotic expression of $\frac{a_n^2}{n}\widehat{A}_{1n}$. Developing the expression of the latter, we will show that  
\begin{align}\label{eq:A1n_hat} 
\frac{a_n^2}{n}\widehat{A}_{1n} = \frac{a_n}{n}\sum_{i=1}^n\psi_{\tau}(\epsilon_i)W_i(\bx)\big[\widehat{c}^{u}_{Y\bX}\big(F^{u}_Y(Y_i), \bF^{u}(\bx)\big) - c^{u}_{Y\bX}\big(F^{u}_Y(Y_i), \bF^{u}(\bx)\big) \big]+o_p\left(1\right).
\end{align}
To that end, note that $\frac{a_n^2}{n}\widehat{A}_{1n}$ may be decomposed as 
$$
\frac{a_n^2}{n}\hat{A}_{1n}=\frac{a_n}{n}\sum_{i=1}^n\psi_{\tau}(\epsilon_i)W_i(\bx)\big[\widehat{c}^{u}_{Y\bX}\big(F^{u}_Y(Y_i), \bF^{u}(\bx)\big) - c^{u}_{Y\bX}\big(F^{u}_Y(Y_i), \bF^{u}(\bx)\big) \big]+\Delta_{3n}+\Delta_{4n}+\Delta_{5n}+\Delta_{6n},
$$
where 
\begin{align}
\Delta_{3n}&=\frac{a_n}{n}\sum_{i=1}^n\psi_{\tau}(\epsilon_i)(\widehat{W}_i(\bx)-W_i(\bx))\widehat{c}^{u}_{Y\bX}\big(\widehat{F}^{u}_Y(Y_i), \widehat{\bF}^{u}(\bx)\big)\label{eq:Delta_3n}\\
\Delta_{4n}&=\frac{a_n}{n}\sum_{i=1}^n\psi_{\tau}(\epsilon_i)W_i(\bx)\big[\widehat{c}^{u}_{Y\bX}\big(\widehat{F}^{u}_Y(Y_i), \widehat{\bF}^{u}(\bx)\big)-\widehat{c}^{u}_{Y\bX}\big(F^{u}_Y(Y_i), \widehat{\bF}^{u}(\bx)\big)\big]\label{eq:Delta_4n}\\
\Delta_{5n}&=\frac{a_n}{n}\sum_{i=1}^n\psi_{\tau}(\epsilon_i)W_i(\bx)\big[\widehat{c}^{u}_{Y\bX}\big(F^{u}_Y(Y_i), \widehat{\bF}^{u}(\bx)\big)-\widehat{c}^{u}_{Y\bX}\big(F^{u}_Y(Y_i), \bF^{u}(\bx)\big)\big]\label{eq:Delta_5n}\\
\Delta_{6n}&=\frac{a_n}{n}\sum_{i=1}^n\psi_{\tau}(\epsilon_i)W_i(\bx) c^{u}_{Y\bX}\big(F^{u}_Y(Y_i), \bF^{u}(\bx)\big)\label{eq:Delta_6n}.
\end{align} 
We will show that all these quantities converge to $0$ in probability. Starting with $\Delta_{3n}$, we have, for a large $n$, 
\begin{align*}
|\Delta_{3n}|&=\left |\frac{a_n}{n}\sum_{i=1}^n\psi_{\tau}(\epsilon_i)W_i(\bx) \widehat{c}^{u}_{Y\bX}\big(\widehat{F}^{u}_Y(Y_i), \widehat{\bF}^{u}(\bx)\big) \frac{\widehat{G}_C(Y_i-|\bx)-G_C(Y_i-|\bx)}{1-\widehat{G}_C(Y_i-|\bx)}\right|\\
& \le a_n \sup_{t\in \RR}\widehat{c}^{u}_{Y\bX}\big(\widehat{F}^{u}_Y(t), \widehat{\bF}^{u}(\bx)\big) \sup_{t\leq \max_{\stackrel{1\leq i\leq n}{\Delta_i=1}} Y_i}\frac{\left|\widehat{G}_C(t|\bx)-G_C(t|\bx)\right|}{1-\widehat{G}_C(t|\bx)} \, \frac{1}{n} \sum_{i=1}^n |\psi_{\tau}(\epsilon_i)|W_i(\bx) \\
&=O_p(a_n) \sup_{t\leq \tau_{F_Y}}\frac{\left|\widehat{G}_C(t|\bx)-G_C(t|\bx)\right|}{1-\widehat{G}_C(t|\bx)}=o_p(1),
\end{align*} 
under the condition that assumptions \ref{cond:pt x}, \ref{cond:techn}-(i), \ref{cond:G_C_hat} and \ref{cond:cop_hat}-(i) are met.\\

The proofs of \eqref{eq:Delta_4n} and \eqref{eq:Delta_5n} are very similar in spirit. Hence, for the sake of brevity, we only consider here \eqref{eq:Delta_4n}. For any $i=1,\ldots,n,$ there exists a $\theta_i\in(0,1)$ such that, 
\begin{multline*}
\widehat{c}^{u}_{Y\bX}\big(\widehat{F}^{u}_Y(Y_i), \widehat{\bF}^{u}(\bx)\big) - \widehat{c}^{u}_{Y\bX}\big(F^{u}_Y(Y_i), \widehat{\bF}^{u}(\bx)\big) = \big(\widehat{F}^{u}_Y(Y_i)-F^{u}_Y(Y_i)\big) \times \\
\partial_1 \widehat{c}^{u}_{Y\bX}\Big(F^{u}_Y(Y_i)+\theta_i\big(\widehat{F}^{u}_Y(Y_i)-F^{u}_Y(Y_i)\big),\,\widehat{\bF}^{u}(\bx)\Big),
\end{multline*}
where $\partial_j$ denotes the partial derivative with respect to the $j$-th argument. Therefore, for a large $n$ and under the bandwidth requirement,
\begin{align*}
|\Delta_{4n}|&\le a_n \sup_{t\in \RR}|\widehat{F}^{u}_Y(t)-F^{u}_Y(t)| \sup_{u_0\,\in\,(0,1)}\sup_{\bm{u}\in\mathcal{V}_{\bF^{u}(\bx)}}\left|\partial_1\,\widehat{c}^{u}_{Y\bX}(u_0,\bm{u})\right|\frac{1}{n}\sum_{i=1}^n|\psi_{\tau}(\epsilon_i)|W_i(\bx)\\
&=o_p(1),
\end{align*}
provided assumptions \ref{cond:techn}-(i), \ref{cond:marg}, and \ref{cond:cop_hat}-(ii) are satisfied. Similarly, $\Delta_{5n}=o_p(1)$.  

Lastly, $\Delta_{6n}=o_p(1)$, since
$$
\E(\psi_{\tau}(\epsilon)\,W(\bx)\,c^{u}_{Y\bX}\big(F^{u}_Y(Y), \bF^{u}(\bx)\big))=0,
$$
and, by assumption \ref{cond:techn}-(ii), 
$$
\frac{1}{n}\sum_{i=1}^n\psi_{\tau}(\epsilon_i)\,W_i(\bx)\,c^{u}_{Y\bX}\big(F^{u}_Y(Y_i), \bF^{u}(\bx)\big)=O_p(n^{-1/2}).
$$
Finally, inserting \eqref{eq:A1n_hat} in \eqref{eq:quadr_lemma}, we conclude that   
\begin{multline*}
a_n(\widehat{m}_{\tau}(\bx)-m_{\tau}(\bx))=\\ \frac{w(\bx)}{f_{T|\bX}(m_{\tau}(\bx)|\bx)} \, \frac{a_n}{n}\sum_{i=1}^n\psi_{\tau}(\epsilon_i)W_i(\bx)\big[\widehat{c}^{u}_{Y\bX}\big(F^{u}_Y(Y_i), \bF^{u}(\bx)\big) - c^{u}_{Y\bX}\big(F^{u}_Y(Y_i), \bF^{u}(\bx)\big) \big] + o_p(1),
\end{multline*}
which completes the proof.
\end{proof}

\subsection*{Proof of Corollary \ref{corr1}}
\begin{proof}[\nopunct]
Given that, by the result of Theorem \ref{theorem1}, the asymptotic behavior of $\widehat{m}_\tau(\bx)$ will be dictated by the expression of the multivariate copula estimator, we will concentrate on the latter. Using the asymptotic expressions of both bivariate copulas along with the multivariate copula decomposition of Section \ref{section:methodology}, under the required bandwidth conditions we have that, 
\begin{multline*}
a_n\left(\widehat{c}^{u}_{Y\bX}\big(u_0, \bm{u}\big) - c^{u}_{Y\bX}\big(u_0, \bm{u}\big) - h^2 b_{Y\bX}(u_0,\bm{u})\right) = \frac{1}{\sqrt{n}}\sum_{j=1}^n\widetilde{Z}^{nj}(u_0,\bm{u})+o_p(1), \\ \forall \bm{u} \in  (0,1)^2, \text{ uniformly in } u_0 \in  (0,1),
\end{multline*}
where 
\begin{align*}
\widetilde{Z}^{nj}(u_0,\bm{u}) &= \left( Z^{nj}_1(u_0,u_1)  c^{u}_{2}\big(u_0, u_2\big) + Z^{nj}_2(u_0,u_2) c^{u}_{1}\big(u_0, u_1\big)\right) c^{u}_{X_1X_2|Y}(u_1,u_2|u_0) \\
b_{Y\bX}(u_0,\bm{u}) &= \left[ b_1(u_0,u_1) c^{u}_{2}\big(u_0, u_2\big) + b_2(u_0,u_2) c^{u}_{1}\big(u_0, u_1\big)  \right] c^{u}_{X_1X_2|Y}(u_1,u_2|u_0).
\end{align*}
Therefore, using the result of Theorem \ref{theorem1}, we may determine that 
\begin{align*}
& a_n\left(\widehat{m}_{\tau}(\bx)-m_{\tau}(\bx) - h^2 \frac{w(\bx)}{f_{T|\bX}(m_{\tau}(\bx)|\bx)} n^{-1} \sum_{i=1}^n \psi_{\tau}(\epsilon_i)W_i(\bx) b_{Y\bX}\left(F^{u}(Y_i),\bF^{u}(\bx)\right) \right)\\
&=\frac{w(\bx)}{f_{T|\bX}(m_{\tau}(\bx)|\bx)}
\sqrt{n}\frac{1}{n^2}\sum_{i=1}^n\sum_{j=1}^n\psi_{\tau}(\epsilon_i)W_i(\bx)\widetilde{Z}^{nj}\left(F^{u}(Y_i),\bF^{u}(\bx)\right)
+o_p(1).
\end{align*}
Let us define
$$
B(\bx)= \frac{w(\bx)}{f_{T|\bX}(m_{\tau}(\bx)|\bx)} \E \left(\psi_{\tau}(\epsilon)W(\bx) b_{Y\bX}\left(F^{u}(Y),\bF^{u}(\bx)\right)\right).
$$
Provided that assumption \ref{cond:techn_biais} holds, we have 
$$
\frac{w(\bx)}{f_{T|\bX}(m_{\tau}(\bx)|\bx)} n^{-1} \sum_{i=1}^n \psi_{\tau}(\epsilon_i)W_i(\bx) b_{Y\bX}\left(F^{u}(Y_i),\bF^{u}(\bx)\right) = B(\bx) + O_p(n^{-1/2}),
$$
from which it results that
\begin{align}\label{eq:corr asy}
& a_n\left(\widehat{m}_{\tau}(\bx)-m_{\tau}(\bx) - h^2 B(\bx)\right)\nonumber\\
&=\frac{w(\bx)}{f_{T|\bX}(m_{\tau}(\bx)|\bx)}
\sqrt{n}\frac{1}{n^2}\sum_{i=1}^n\sum_{j=1}^n\psi_{\tau}(\epsilon_i)W_i(\bx)\widetilde{Z}^{nj}\left(F^{u}(Y_i),\bF^{u}(\bx)\right)
+o_p(1).
\end{align}
The last step of the proof is then to simplify the obtained expression on the right hand side of the last equality. To that end, define $V_n = n^{-2}\sum_{i=1}^n\sum_{j=1}^n\psi_{\tau}(\epsilon_i)W_i(\bx)\widetilde{Z}^{nj}\left(F^{u}(Y_i),\bF^{u}(\bx)\right)$, and observe that this is a V-statistic with the symmetric kernel
$$
\Psi_n(V_i,V_j) = \frac{1}{2}\left[\psi_{\tau}(\epsilon_i)W_i(\bx)\widetilde{Z}^{nj}\left(F^{u}(Y_i),\bF^{u}(\bx)\right) + \psi_{\tau}(\epsilon_j)W_j(\bx)\widetilde{Z}^{ni}\left(F^{u}(Y_j),\bF^{u}(\bx)\right)\right],
$$
where $V_t = (Y_t,\bX_t,\Delta_t), t=i,j$. As the statistic's kernel depends on $n$, this suggests to apply Corollary 1 in \citet{MY06} which establishes the $\sqrt{n}$-equivalence between the V-statistic and the H\'{a}jek-projection of its corresponding U-statistic (see e.g. \citet{S80}, page 189). Therefore, as $\E\left(Z^{nj}_{k}(u_0,u_k)\right)=0$ for all $u_0,u_k$, $k=1,2,$ and $\E\left(\Psi_n(V_i,V_j)\right)=0$, we have
\begin{align}\label{eq:corr V stat}
V_n = n^{-1} \sum_{i=1}^n  \lambda_n\left(Y_i,\Delta_i,\bX_i,\bx\right) + o_p(n^{-1/2}),
\end{align}
where $\lambda_n\left(Y_i,\Delta_i,\bX_i,\bx\right)=\E\left[\psi_{\tau}(\epsilon)W(\bx)\widetilde{Z}^{ni}\left(F^{u}(Y),\bF^{u}(\bx)\right)|Y_i,\Delta_i,\bX_i\right]$, provided that assumption \ref{cond:techn_Vstat} holds. Lastly, the result of Corollary \ref{corr1} follows readily from the insertion of \eqref{eq:corr V stat} in \eqref{eq:corr asy}  and the application of Lyapunov's central limit theorem.
\end{proof}

\bibliographystyle{plainnat}
\bibliography{CQRCD}

\end{document}